\documentclass[english,letterpaper]{article}
\usepackage[T1]{fontenc}
\usepackage[utf8]{inputenc}
\usepackage{array}
\usepackage{booktabs}
\usepackage{units}
\usepackage{multirow}
\usepackage{amsmath}
\usepackage{amsthm}
\usepackage{amssymb}
\usepackage[authoryear]{natbib}

\makeatletter

\providecommand{\tabularnewline}{\\}

\relax
\usepackage{general22}  

\usepackage{times}  
\usepackage{helvet}  
\usepackage{courier}  
\usepackage[hyphens]{url}  
\usepackage{graphicx} 
\usepackage{natbib}  
\usepackage{caption} 
\DeclareCaptionStyle{ruled}{labelfont=normalfont,labelsep=colon,strut=off} 
\usepackage{algorithm}
\usepackage{algorithmic}

\usepackage{newfloat}
\usepackage{listings}
\floatstyle{ruled}
\newfloat{listing}{tb}{lst}{}
\floatname{listing}{Listing}

\usepackage{soul}
\usepackage{url}
\usepackage{doi}

\usepackage{hyperref}
\usepackage{xcolor}
\hypersetup{%
  colorlinks=false,
  linkbordercolor=black,
  urlbordercolor=black,
  citebordercolor=black,
  pdfborderstyle={/S/U/W 0}
}

\usepackage[utf8]{inputenc}
\usepackage{amsthm}
\usepackage{booktabs}
\newcommand{\yes}{\oldstylenums{1}}
\newcommand{\no}{\oldstylenums{-1}}
\newcommand{\zero}{\oldstylenums{0}}

\usepackage{balance} 
\usepackage{textpos} 

\usepackage{tikz}
\usepackage{pgfplots}

\renewcommand{\tfrac}[2]{\genfrac{}{}{0pt}{2}{#1}{#2}}
\usepackage{colortbl}
\newcommand{\colagt}{\cellcolor{blue!10}}

\title{Detecting Coordinated Inauthentic Behavior in Likes on Social Media:\\ Proof of Concept
}
\author {
    Affiliations blinded for Review 
}

\usepackage{titling}
\usepackage{authblk}

\AtBeginDocument{
  
}

\makeatother

\usepackage{babel}
\begin{document}
\pagenumbering{arabic} 
\setcounter{page}{1}
\author[1]{Laura Jahn}
\author[1]{Rasmus K. Rendsvig}
\author[1,2]{Jacob St{\ae}rk-{\O}stergaard}
\affil[1]{Center for Information and Bubble Studies, Department of Communication, University of Copenhagen}
\affil[2]{Animal Welfare and Disease Control, Department of Veterinary and Animal Science, University of Copenhagen}
\date{}

\maketitle
\begin{abstract}
Coordinated inauthentic behavior is used as a tool on social media
to shape public opinion by elevating or suppressing topics using systematic
engagements—e.g. through `likes' or similar reactions. In an honest
world, reactions may be informative to users when selecting on what
to spend their attention: through the wisdom of crowds, summed reactions
may help identifying relevant and high-quality content. This is nullified
by coordinated inauthentic liking. To restore wisdom-of-crowds effects,
it is therefore desirable to separate the inauthentic agents from
the wise crowd, and use only the latter as a voting `jury' on the
relevance of a post. To this end, we design two \emph{jury selection
procedures} (\textsc{jsp}s) that discard agents classified as inauthentic.
Using machine learning techniques, both cluster on binary vote data—one
using a Gaussian Mixture Model (\textsc{gmm} \textsc{jsp}), one the
$k$-means algorithm (\textsc{km} \textsc{jsp})—and label agents by
logistic regression. We evaluate the jury selection procedures with
an agent-based model, and show that the \textsc{gmm jsp} detects more
inauthentic agents, but both \textsc{jsp}s select juries with vastly
increased correctness of vote by majority. This proof of concept provides
an argument for the release of reactions data from social media platforms
through a direct use-case in the fight against online misinformation.\medskip{}

\textbf{Keywords}: Coordinated inauthentic behavior, bot detection,
social media, wisdom of crowds, simulation, agent-based modeling
\end{abstract}

\section{Introduction}

In April 2022, we bought 100 Twitter likes for 3.85 USD through a
readily accessible website. These 100 likes sufficed to catapult the
liked tweet to the top of the \emph{Top} feed of \texttt{\#dkpol},
the main Twittersphere for discussing Danish politics. There, it stayed
for several hours.\footnote{When the hashtag was viewed in private browser tab without being logged
or when logged in with a new Twitter profile. The tweet was clearly
marked as a test, and published by an account with almost no network
or activity.} This illustrates that Twitter's content sorting algorithm may be
easily hacked to bring selected items to users' attention using only
likes.

Our tweet was clearly marked as off-topic for \texttt{\#dkpol}, but
could have been misinformation. Our ``inauthentic likes'' could
thus have been used with the intent to mislead or manipulate—and this
would not be uncommon: when deploying \emph{influence operations}
(IOs) on social media platforms\emph{ }to shape public opinion\emph{
}\citep{nizzoli2021coordinated}, a central strategy is to exploit
the platforms' content sorting algorithms to highlight posts to users,
a process known as \emph{attention hacking }\citep{Goerzen2019}.
Attention hacking through likes requires coordination of likes to
maximize effect. As the liking behavior does not reflect authentic
personal beliefs, it is an example of so-called \emph{coordinated
inauthentic behavior }(CIB) \citep{Pacheco2021Coordinated,Schoch2022,nizzoli2021coordinated}.\footnote{As many platforms' sorting algorithms assign higher rank to posts
that many users have engaged with—e.g., through liking, upvoting,
sharing, retweeting or commenting—attention hacking influence operations
orchestrate coordinated engagements through coordinated inauthentic
behavior to maximize their effect \citep{nizzoli2021coordinated}.} Coordinated inauthentic behavior may be exhibited by humans and bots
alike.

Liking is an engagement type common across social media platforms,
but as different platforms use different labels, we refer to \emph{reactions},
understood as one-click engagements where users may select one option
from a short pre-defined list as their `reaction' to a post, with
users' choices typically summed and presented as a quantified metric
beneath the item. Reactions include perhaps most famously Facebook's
original `Like' and their now five other reaction emojis, the hearts/likes
on Instagram, TikTok and Twitter, and Reddit's up- and downvotes \citet{weber2021amplifyingcoord}.
Importantly, all these reactions inform the platforms' algorithmic
content sorting, and thus steer users' attention.

In an honest world, reactions may be informative in steering attention:
through the wisdom-of-crowds, summed reactions may help identify relevant,
well-produced, or otherwise high quality content as attention-worthy,
so it may be presented to users at the top of their news feed \citep{Bhadani2022_Nature_diversity}.
Alas, that reactions serve as attention-steering exactly makes them—along
with other quantified attention metrics \citep{Giglietto2020coordinated}—a
target candidate for influence operations that spread misinformation
based on coordinated inauthentic behavior (CIB-based IOs). Accounts
(often bots) used to hack users' attention simulate authentic interest
in a topic through reacting to social media posts \citep{Goerzen2019}.
While not actively posting content, they seek to elevate or suppress
specific topics in the public perception, flood platforms with misinformation,
and boost narratives counter to an authentic public interest \citep{Takacs2019}.
The identification of such computational propaganda is difficult as
modern bots mask their identity, mimicking human behavior to an increased
extent \citep{Beatson2021,Bradshaw2017}.

When CIB-based IOs target reactions, the wisdom-of-crowds effect is
lost. Scholars have called for ways to promote the Internet’s potential
to strengthen rather than diminish democratic virtues \citep{Lazer2018},
e.g., by redesigning online environments to enable informed choice
of attention expenditure by providing transparent crowd-sourced voting
systems \citep{Lorenz-Spreen2020}. Here, current implementations
of reactions are in the ballpark, yet strongly flawed as they may
be hacked by CIB-based IOs. Adopting exactly a voting perspective,
this paper develops a computational approach to detect and remove
CIB influence on reactions, with the aim to restore reactions' wisdom-of-crowds
effects.

Detecting and removing coordinated inauthentic behavior targeted to
reactions is a neglected area of research (perhaps partially because
relevant data is difficult for researchers to obtain despite often
being public, a topic we return to below and in the concluding remarks).
In general, computational approaches to combat CIB have not been studied
extensively \citep{nizzoli2021coordinated}. Recent research has explored
user information-based coordination such as account handle sharing,
content-based coordination (e.g., synchronized co-posting of images,
hashtags, text, and links), attention metric-based coordination such
as co-retweeting, or timing-based coordination \citep{Kirn2022,Pacheco2021Coordinated,nizzoli2021coordinated,Giglietto2020coordinated,Giglietto2020coordinated2,Grimme2018CoordUnsupervOwnBots,weber2021amplifyingcoord}.
Despite reactions being a commonly adopted and an easily manipulatable
mechanism, research on CIB more narrowly targeted at reactions is
quite scarce. Borderlining relevancy are studies on purchased likes
not of posts, but of \emph{pages} \emph{{[}followers{]}} on Facebook
{[}Twitter{]} \citep{ikram2017_fblikefarms,Cristofaro2014_fblikes,beutel2013copycatch}
{[}\citep{aggarwal2015Twitterfakefol}{]}. This stream of work tries
to understand the modus operandi of page like farms {[}follower farms{]}
\citep{Cristofaro2014_fblikes} {[}\citep{aggarwal2015Twitterfakefol}{]}
and develops supervised classification models based on demographic,
temporal, and social characteristics \citep{ikram2017_fblikefarms}
{[}\citep{aggarwal2015Twitterfakefol}{]}. Here, notably, \citet{ikram2017_fblikefarms}
find that their bot classifier has difficulty detecting page like
farms that mimick regular like-spreading over longer timespans, and
conclude that \citet{beutel2013copycatch}'s unsupervised approach
to detect page like farms—even developed with data from inside Facebook—yielded
large false positive errors.\footnote{Also in the closely related field of bot detection has the detection
of bots that are mainly designed to engage through reactions gone
unstudied, again perhaps due to data restrictions. For a systematic
review of the bot detection literature, see \citep{Orabi2020}.} Directly about reactions to posts is \citet{Torres-Lugo_Likes_Manip_deletions}'s
study of metric inflation through strategic deletions on Twitter.
They analyze coordination in repetitive \emph{(un)liking} on \emph{deleted}
tweets in influence operations that seek to bypass daily anti-flooding
tweeting limits. From a curation point of view, looking at unlikes
is a very smart move, as this data is in fact available to purchase
from Twitter. Alas, the approach is inapplicable to tweets that remain
online, such as those central to CIB-based IOs that push narrative
through political astroturfing \citep{Schoch2022}.

\subsection{A Voting and Simulation Approach to Coordinated Inauthentic Behavior}

To study CIB targeted at reactions, we methodologically take a voting
perspective on reactions and a computer simulation approach to validate
the proposed methods.

With the voting perspective, we conceptualize reactions as votes about
the epistemic quality of an information item. We restrict attention
to a two-reaction case, with one reaction interpreted as a vote \emph{for}
the item being of high quality, the other a vote against. We adopt
this voting perspective as it allows us to clearly explicate a structure
of reactions as binary voting, to specify different patterns and varying
degrees of coordination \citep{nizzoli2021coordinated}, and to define
and quantify the aptitude of a group of users with respect to tracking
quality.

Further, it allows us to draw on intuitions from the \emph{Condorcet
Jury Theorem}\footnote{When all jurors vote \emph{independently} and are \emph{better than
random} at voting correctly, the probability of a correct majority
judgment approaches $1$ as the jury size approaches~$\infty$.} \citep{Condorcet}: while many weakly competent authentic judgments
may lead to a highly accurate collective judgment through simple majority
vote, such positive wisdom-of-crowds effects may be counteracted by
the non-independence exhibited by coordinated inauthentic behavior.

The latter motivates the paper's fundamental approach to counter CIB
influence, namely to design \emph{jury selection procedures} (\textsc{jsp}s).
The core idea is this: given a collection of votes from a voting population
of agents, a \textsc{jsp} searches the collection for coordinated
voting and from the findings classifies agents as inauthentic or authentic,
before finally returning a subset of the population—the \emph{jury—}whose
votes are tallied to determine the epistemic quality of a post. I.e.,
a \textsc{jsp }censors a subset of the population's votes in order
to restore wisdom-of-crowds effects for the remainder.

Methodologically, the paper is also a computer simulation paper. We
develop an agent-based model (ABM) in which agents vote on the quality
of fictitious posts. The ABM includes agents that vote authentically—in
accordance with their private beliefs about the quality of the post
and the assumptions of the Condorcet Jury Theorem—and some that do
not, either by voting only inauthentically or coordinatedly inauthentically.
Over synthetic vote data generated by the ABM, we test and validate
the machine learning-based \textsc{jsp}s that we develop.

Validating with synthetic data circumvents three main challenges in
detecting coordinated inauthentic users (lacking reproducibility,
lacking data availability, and lacking ground truth), while suffering
the downside that synthetic data has limited ecological validity.
First, empirical social media studies of bots remain problematic to
replicate and reproduce due to a time-sensitivity of the relevant
data \citep{Martini2021,samper2021bot,bebensee2021leveraging}. Attempts
to collect the same data twice are likely to fail, as traces of coordination
may be altered or deleted after an influence operation was concluded.
While e.g. Twitter grants generous academic research access to historic
tweets through their API, accounts involved in CIB may evade detection
as they are no longer retrievable in their original appearance \citep{Torres-Lugo_Likes_Manip_deletions}.
The shortcomings in data reproducibility make CIB/bot detection frameworks
difficult to compare, as these typically require live data access
\citep{Martini2021}. Data and analyses of the methods proposed here
are time-insensitive and reproducible (cf. Data Availability Statement
and Supplementary Material\footnote{Code to reproduce and analyse the data can be found at the GitHub repository \href{https://github.com/LJ-9/Coordinated-Inauthentic-Behavior-Likes-ABM-Analysis}{LJ-9/Coordinated-Inauthentic-Behavior-Likes-ABM-Analysis}.}.

Second, data availability limits research. Large scale studies may
simply be impossible due to data access restrictions \citep{Martini2021,Disinfresearch-agenda-2020,misinfo_data}.
Specifically data concerning users' reactions is very difficult for
researchers to obtain: none of the currently existing datasets include
it,\footnote{See e.g. Indiana University's Bot Repository, a resourceful, centralized
repository of annotated datasets of Twitter social bots: https://botometer.osome.iu.edu/bot-repository/datasets.html.} and neither Meta, Twitter nor Reddit supply this data in necessary
scope \citep{Disinfresearch-agenda-2020,misinfo_data}. We outline
data collection strategies in connection with empirical validation
of our methods in the concluding remarks. Data from an ABM can be
(re)synthesized in any quantity.

Third, there is an issue with lacking ground truth as researchers
do not have access to the empirical truth about accounts involved
in coordinated inauthentic behavior. Qualified guesses can be made
based on suspicious similarities in behavior or profile features,
but \emph{de facto}, it remains unknown whether two users' actions
are authentically correlated or inauthentically coordinated, or how
many fully or partially automated accounts exist in a total population
\citep{Magelinski2022,Martini2021,samper2021bot,Chavoshi2017_unsupervised,beutel2013copycatch}.
Specifically for reaction-based CIB, it seems infeasible to create
a labeled dataset that even \emph{approximates} the ground truth:
labeling accounts individually e.g. via crowd-sourcing or the well-established
bot classifier \emph{Botometer} will likely fail as $i)$ single accounts
will often seem inconspicuous unless looked at in concert at a collective
level \citep{Magelinski2022,Grimme2018CoordUnsupervOwnBots,Yang2019,Yang2020},\footnote{\emph{Botometer}'s feature-based approach considers accounts one at
a time and does therefore not pick up on group anomalies based on
suspicious similarity \citep{Yang2019,Yang2020}.} and $ii)$ collective level labeling is impossible due to current
data restrictions as reactions data is available only in severely
limited quantities, if at all.\footnote{Twitter is the only platform that offers \emph{any }access, with limitations
of 75 requests per 15 minutes, each granting only the most recent
100 liking users of a single tweet. See https://developer.twitter.com/en/docs/twitter-api/tweets/likes/api-reference/get-tweets-id-liking\_users.}

By validating over an ABM where we specify which agents are involved
in CIB, we gain transparency and a ground truth. We get precise baselines,
exact measurements of the effect of our methods, and certainty about
the degrees of misclassification. We elaborate on this below. Hereby,
the ABM validation allows us to provide methodologically robust proof
of concept for the \textsc{jsp} approach. 

\subsection{Existing Work and Contributions}

Little work exists on identifying and eliminating inauthentic votes
and\textsc{ jsp}s. \citet{GaleazziRendsvigSlavkovik2019} suggest
to remove inauthentic influence by identifying an independent jury
via the $\chi^{2}$ test of independence. Their model takes sharing-induced
diffusion in social networks as evolving crowdvoting. Their main results
pertain to \textsc{jsp} time-complexity, with their least requiring
suggestion still exponential in the jury size (a direct consequence
of using $\chi^{2}$). In addition, we find that the number of data
points required for $\chi^{2}$ application (see Sec. \ref{sec:Classification})
makes their \textsc{jsp}s practically inapplicable and computationally
unservicable. A performance comparison with their bot detecting scheme
is therefore impossible beyond contrasting data requirements.

The central goal of this paper is to develop jury selection procedures
that raise the correctness of vote by majority of juries, complementing
\citep{GaleazziRendsvigSlavkovik2019}. The methodological voting
perspective allows us to define a metric of success for the methods
we develop: majority correctness scores (\textsc{mcs}s). Majority
correctness scores give a direct perspective on the collective epistemic
practice of a group of agents, providing a more conclusive perspective
than misclassification scores. Beyond raising majority correctness
scores, we desire \emph{accurate} \textsc{jsp}s that minimize misclassification
of $i.$ authentic agents as inauthentic and $ii.$ inauthentic agents
as authentic (i.e., minimize $i.$ false positive and $ii.$ false
negative errors). The first values \emph{vox populi} and penalizes
censorship \citep{Shao_Menczer_NatComm2018}, while the second is
a \emph{precautionary principle} against inauthentic influence. Further,
we desire \emph{feasible} \textsc{jsp}s that use only data that is
obtainable by social media platforms and that requires little to no
preprocessing, have few to no supervised elements \citep{Orabi2020,Grimme2018CoordUnsupervOwnBots},
and have reasonable complexity.

This paper develops two \textsc{jsp}s, evaluated with respect to vote
data generated by the agent-based model. The ABM is presented in Sec.~\ref{sec:Agent-Based-Model-(ABM)}
where varying baseline agent populations' majority correctness scores
\textsc{(mcs}s) are inspected, on which inauthentic activity has a
substantial negative impact.

The core \textsc{jsp} machinery is presented in Sec.~\ref{sec:Classification}.
Each jury selection procedure invokes a classifier method that decomposes
the ABM data into singular values (SVD), applies a clustering strategy
(either a Gaussian Mixture Model (\textsc{gmm}) or the $k$-means
algorithm (\textsc{km})), and labels agents using a non-standard application
of logistic regression on the qualitative property of the post voted
on. In related work, vote data—such as US congress roll call data—has
been successfully grouped employing dimensionality reduction, e.g.,
\citep{Yang2020_1dim_congress,Porter2005,Sirovich2003,Poole2000}.
Our approach is novel in applying such methods in the realm of digital
propaganda using simple binary input data. Dimensionality reduction
and clustering methods have so far been applied to less sparse data
structures, such as HTTP-level traffic patterns \citep{Suchacka2019,Suchacka2020},
textual data of tweets \citep{Kirn2022}, or rich datasets with behavior-based
features (number of friends/followers, mentions and hashtags, etc.)
like in detection of spam bots on social media sites (e.g., \citep{Ahmed2013}).
A systematic review on detection of bots on social media \citep{Orabi2020}
further discusses unsupervised methods, e.g. \citep{Chavoshi2017_unsupervised,Chen_Unsupervised},
yet to our knowledge only \citet{GaleazziRendsvigSlavkovik2019} attempt
to flag agents given just binary vote data (i.e., with no added information
about e.g. temporal coordination as in \citep{beutel2013copycatch,Grimme2018CoordUnsupervOwnBots,Magelinski2022,Pacheco2021Coordinated,Schoch2022})
obtainable intra-platform by social media sites.

In Sec.~\ref{sec:Jury-Selection-and}, we define and evaluate the
\textsc{gmm} and \textsc{km} jury selection procedures. We show that
both are highly successful, as they select juries that have vastly
increased majority correctness scores compared to baseline juries.
Moreover, the \textsc{gmm} \textsc{jsp} outperforms the \textsc{km}
\textsc{jsp} with respect to its accurate and particularly precautious
results. Sec.~\ref{sec:Concluding-Remarks} summarizes the main findings
and discusses ethical considerations, model assumptions, and data
collection.

Technically, we contribute a novel, reactions-based approach to detect
CIB, implemented in two variants evaluated to have positive effects
over synthetic ABM data, thus showing proof of concept. Societally,
the proof of concept provides a direct argument to be raised to social
media platform to open access to reactions data: the data is necessary
to evaluate, tweak and deploy promising methods (i.e., \textsc{jsp}s)
to combat coordinated inauthentic behavior and thus to inhibit the
spread of misinformation.

\section{\label{sec:Agent-Based-Model-(ABM)}Agent-Based Model (ABM)}

We evaluate the two jury selection procedures over data generated
by the following agent-based model. A model\emph{ run} consists of
a fixed set of agents partitioned into agent types (see below), and
a sequence of independent \emph{voting rounds}. Each round concerns
a given post (which we do not explicitly represent) and whether the
post is of high or low quality, on which agents vote $\{\yes,\no\}$
($\yes$ for high, $\no$ for low). We think of these votes as users'
reactions, and call $\yes$ an \emph{upvote} and $\no$ a \emph{downvote}.

Agents are either \emph{authentic} or \emph{inauthentic}. We formally
define the agents types in Sec.~\ref{subsec:Agent-Types} below.
We think of authentic agents as regular social media users that use
their up- and downvotes to inform about post quality (e.g., analogously
to \citet{Metaxas2015} who showed that by retweeting, users on Twitter
signal trust in the message). Authentic agents vote independently
according only to their competence-based beliefs about post quality:
they satisfy the assumptions of the Condorcet Jury Theorem. Inauthentic
agents do not: with different patterns and varying degrees, they coordinate
their votes through properties distinct from quality. On social media,
inauthentic behavior can both be witnessed among human controlled
and automated accounts. Given the scale of influence operations, it
is relevant to think about inauthentic behavior in terms of so-called
\emph{social bots}: ``\emph{Computer programs designed to use social
networks by simulating how humans communicate and interact with each
other}” \citep{Abokhodair2015}. The design of our inauthentic agents
draws inspiration from the social bot classes\emph{ astroturfing bots}
(that create ``\emph{the appearance of widespread support for a candidate
or opinion}” \citep{Ratkiewicz2011}) and \emph{influence bots} (“\emph{Realistic
automated identities that illicitly shape discussion}” \citep{Subrahmanian2016})
\citep{Orabi2020}.

\subsection{Post Properties, Competences and Beliefs}

Let $A$ be a finite set of agents and $I=\{1,2,3\}$ an index set
for properties. A voting round concerns a given post, and commences
with the (Monte Carlo like) sampling of a state 
\[
s=(p_{i},C_{a}(p_{1}),B_{a}(p_{i}))_{i\in I\cup A,a\in A}\in\mathbb{R}^{3+|A|+|A|+3|A|+|A|^{2}}
\]
where each $p_{i}$ represents a property of the post, $C_{a}(p_{1})$
is agent $a$'s competence in evaluating whether the post has property
$p_{1}$ and $B_{a}(p_{i})$ is $a$'s belief about whether the post\textsc{
}has property $p_{i}$. Properties $(p_{i})_{i\in I}=(p_{1},p_{2},p_{3})\in\{\no,\yes\}^{3}$
are sampled independently from a binomial distribution with probabilities
$P(p_{i}=\yes)=(1-P(p_{i}=\no))$, given as noise levels in Sec.~\ref{subsec:Parameters-and-Generated-Dataset}.
Each $p_{a}$ is sampled as $p_{3}$, and is a private property used
by some agent types.\footnote{We include $p_{a}$ and $B_{a}(p_{b})$ for all agents $a,b\in A$
in the state for description simplicity. In the simulation implementation,
we only sampled $p_{a}$ and $B_{a}(p_{a})$ for agents $a$ that
make use of $p_{a}$.} We say that the post has property $p_{i}$ if $p_{i}=\yes$, else
that it does not. Property $p_{1}$ represents whether the post has
high or low quality, and is the only property relevant to authentic
agents. Inauthentic agents act also on additional properties, as described
below.

Each agent $a\in A$ is assigned a competence $C_{a}(p_{1})$ to determine
whether the post has high quality, $p_{1}$.\footnote{Even if $p_{1}$ is irrelevant to the agent's voting behavior. This
is to simplify the implementation of the model simulation.\label{fn:1}} To evaluate $p_{1}$, it is assumed that all agents are better than
fair coin tosses but not perfect: $C_{a}(p_{1})\in[0.65,0.95]$. We
chose $[0.65,0.95]$ for $C_{a}(p_{1})$ to expedite convergence towards
a 100\% \textsc{mcs} for authentic agents while ensuring imperfect
competence. Any closed, convex subinterval of the open $(0.5,1)$
would yield similar results w.r.t. \textsc{mcs}, more or less quickly.
Competences are uniformly resampled each round, to capture that agents'
expertise may vary from post to post. Inauthentic agents are assumed
perfectly competent in evaluating properties $p_{2}$ and $p_{3}$,
which they use to coordinate their actions: $C_{a}(p_{2})=C_{a}(p_{3})=1$.
Properties and competences probabilistically determine agents' beliefs:
for all $a\in A,i\in I$, the beliefs $B_{a}(p_{i})\in\{\no,\yes\}$
are sampled with 
\begin{equation}
C_{a}(p_{i})=P(B_{a}(p_{i})=p_{i}).\label{eq:beliefs}
\end{equation}
\setcounter{footnote}{0} If $B_{a}(p_{i})=\yes$, then $a$ believes
that the post has property $p_{i}$, else $a$ believes it does not.\footnote{Hence, agents never suspend judgment, even on properties irrelevant
to their voting behavior. Superfluous beliefs have no effects, and
are only to simplify implementation.} If $B_{a}(p_{i})=p_{i}$, then $a$'s belief about $p_{i}$ is correct.
Hence, (\ref{eq:beliefs}) states that the probability of agent $a$'s
beliefs about $p_{i}$ being correct equates $a$'s competence with
respect to $p_{i}$. For two rounds and their states $s$ and $s'$,
all sampling is independent, and in each state $s$, each $C_{a}(p_{1})$
is independent from $C_{b}(p_{1})$, $a\ne b$. No correlations between
properties are assumed due to the interpretations of $p_{2}$ and
$p_{3}$, stated below. 

\subsection{Agent Types\label{subsec:Agent-Types}}

We define 10 agent types. Each agent type is a behavior-defining function
that maps an agent's beliefs to votes. The set of agent types is $\{\mathtt{A},\mathtt{B}_{i},\mathtt{D}_{i},\mathtt{L}_{i}\}_{i\in\{\uparrow,\downarrow,\updownarrow\}}$,
each defined and described below. A \emph{population }is a map $\mathcal{P}:A\longrightarrow\{\mathtt{A},\mathtt{B}_{i},\mathtt{D}_{i},\mathtt{L}_{i}\}_{i\in\{\uparrow,\downarrow,\updownarrow\}}$
that assigns each agent an agent type.

Intuitively, $\{\mathtt{A},\mathtt{B}_{i},\mathtt{D}_{i},\mathtt{L}_{i}\}_{i\in\{\uparrow,\downarrow,\updownarrow\}}$
contains the following agent types: $\mathtt{A}$ is the \emph{authentic}
agent type, and the inauthentic agents come in three types that incorporate
different patterns of coordination—\emph{boosters} $\mathtt{B}_{i}$,
\emph{distorters} $\mathtt{D_{i}}$, and \emph{lone wolfs} $\mathtt{L}_{i}$.
Each inauthentic type votes based on beliefs about a property \emph{distinct}
from quality. Boosters and distorters vote respectively given properties
$p_{2}$ and $p_{3}$ to coordinate their inauthentic behavior in-group.
Lone wolfs do not coordinate. Each group contains three sub-types:
one main to our story which \emph{upvotes on cue} $(i=\,\uparrow)$,
and two auxiliary that \emph{downvote on cue }$(i=\,\downarrow)$
or \emph{both up-and downvote on cue} $(i=\,\updownarrow)$. We include
the auxiliary sub-types to create a more noisy—and thus harder to
maneuver—setting for the \textsc{jsp}s. We hope the notation is mnemonically
helpful rather than distracting.

Throughout, the largest population is $\mathcal{P}_{\mathbf{\mathtt{Full}}}$,
defined for an agent set $A$, $|A|=1900$, with $1000$ agents assigned
to $\mathtt{A}$ and $100$ agents to each $\mathtt{X}\in\{\mathtt{B}_{i},\mathtt{D}_{i},\mathtt{L}_{i}\}_{i\in\{\uparrow,\downarrow,\updownarrow\}}$.
This size and ratio allows for flexibly choosing subpopulations with
sizes large enough to produce robust votes. We mainly study subpopulations
(restrictions) of $\mathcal{P}_{\mathbf{\mathtt{Full}}}$. We specify
these subpopulations by stating the size of the pre-image of the agent
types (which is sufficient as precise agent identity will not matter),
where we write $|\mathtt{X}|$ for $|\mathcal{P}_{\mathbf{\mathtt{Full}}}\,^{-1}(\mathtt{X})|$
for agent type $\mathtt{X}$. The four main subpopulations are subsets
of either 1000 agents ($\mathcal{P}_{\mathtt{All}}$ containing all
agents types, with 100 agents of each type) or 200 agents ($\mathcal{P_{\mathtt{B}_{\mathrm{\uparrow}}}}$,
$\mathcal{P_{\mathtt{D}_{\mathrm{\uparrow}}}}$ and $\mathcal{P_{\mathtt{L}_{\mathrm{\uparrow}}}}$
each with 100 authentic agents and 100 agents of either type $\mathtt{B}_{\uparrow}$,
$\mathtt{D}_{\uparrow}$ or $\mathtt{L}_{\uparrow}$). Thus, let $\mathcal{P}_{\mathtt{All}}$
be the restriction of $\mathcal{P}_{\mathbf{\mathtt{Full}}}$ with
$|\mathtt{X}|=100$ for each $\mathtt{X}\in\{\mathtt{A},\mathtt{B}_{i},\mathtt{D}_{i},\mathtt{L}_{i}\}_{i\in\{\uparrow,\downarrow,\updownarrow\}}$,
let $\mathcal{P_{\mathtt{B}_{\mathrm{\uparrow}}}}$ be the restriction
of $\mathcal{P}_{\mathbf{\mathtt{Full}}}$ with $|\mathtt{A}|=|\mathtt{B}_{\uparrow}|=100$
and $|\mathtt{X}|=0$ for $\mathtt{X}\in\{\mathtt{B}_{i},\mathtt{D}_{i},\mathtt{L}_{i}\}_{i\in\{\uparrow,\downarrow,\updownarrow\}}\backslash\{\mathtt{B}_{\uparrow}\}$,
and let $\mathcal{P_{\mathtt{D}_{\uparrow}}}$ and $\mathcal{P}_{\mathtt{L_{\uparrow}}}$
be given as $\mathcal{P_{\mathtt{B}_{\mathrm{\uparrow}}}}$ replacing
$\mathtt{B}_{\uparrow}$ with respectively $\mathtt{D}_{\uparrow}$
and $\mathtt{L}_{\uparrow}$. We may further specify subpopulations
of $\mathcal{P}_{\mathbf{\mathtt{All}}}$, $\mathcal{P_{\mathtt{B}_{\mathrm{\uparrow}}}}$,
$\mathcal{P_{\mathtt{D}_{\uparrow}}}$ and $\mathcal{P}_{\mathtt{L_{\uparrow}}}$
like we specify subpopulations of $\mathcal{P}_{\mathbf{\mathtt{Full}}}$.
These restrictions mainly serve to describe what happens when we reduce
the number of authentic agents. We write e.g. ``$\mathcal{P_{\mathtt{B}_{\mathrm{\uparrow}}}}$
for $|\mathtt{A}|=25$'' to mean the subpopulations of $\mathcal{P_{\mathtt{B}_{\mathrm{\uparrow}}}}$
with $125$ agents in total, $25$ of them authentic.

\subsubsection*{Authentic Agents.}

Authentic agents—agents $a$ of type $\mathtt{A}$—correspond to those
assumed in the Condorcet Jury Theorem: they vote fully in accordance
with their beliefs about quality ($p_{1}$), independently of others,
and with a competence strictly above $0.5$. The vote of an authentic
agent $a$ in state $s$ is $\mathtt{A}(a,s)\in\{\no,\yes\}$, given
by the following table:
\begin{center}
\begin{tabular}{c|c|c}
 & $B_{a}(p_{1})=\yes$ & $B_{a}(p_{1})=\no$\tabularnewline
\hline 
$\mathtt{A}(a,s)$ & $\yes$ & $\no$\tabularnewline
\end{tabular}
\par\end{center}

\noindent In this and the below tables, row index ($\mathtt{A}(a,s)$)
denotes the agent type and the cell content denotes the action taken
in the circumstances specified in the column index (e.g. $B_{a}(p_{1})=\yes$).

\subsubsection*{Boosters.}

Boosters vote in a coordinated partisan fashion, aiming to swing the
majority vote in a direction given by $p_{2}$, irrespective of quality
($p_{1}$). Hence boosters exhibit CIB. In social media terms, we
think of $p_{2}$ as disconnected from quality ($p_{1}$), but as
representing that the post\textsc{,} e.g., originates from a specific
source, expresses a given viewpoint, or—taking booster agents as bots—as
tagged for special action by a handler.

The main \emph{Upvote Booster} $\mathtt{B}_{\uparrow}$ has as goal
to boost and amplify $p_{2}$ posts: they upvote (``Yes, the post
has $p_{1}$'') if they believe the post has property $p_{2}$, and
else vote authentically (to hide their inauthentic activities). For
auxiliaries, the \emph{Downvote Booster} $\mathtt{B}_{\downarrow}$
`inverts' $\mathtt{B}_{\uparrow}$: $\mathtt{B}_{\downarrow}$ demotes
non-$p_{2}$ posts by downvoting if they believe the post does not
have $p_{2}$, and else vote authentically, while the \emph{Both Booster}
$\mathtt{B}_{\updownarrow}$ combine the inauthentic behaviors of
$\mathtt{B}_{\uparrow}$ and $\mathtt{B}_{\downarrow}$ by always
voting according to $p_{2}$, and never authentically. The vote of
an agent $a$ of type $\mathtt{B}_{i\in\{\uparrow,\downarrow,\updownarrow\}}$
in state $s$ is $\mathtt{B}_{i}(a,s)$ given by
\noindent \begin{center}
\begin{tabular}{c|c|c}
 & $B_{a}(p_{2})=\yes$ & $B_{a}(p_{2})=\no$\tabularnewline
\hline 
$\mathtt{B}_{\uparrow}(a,s)$ & $\yes$ & $\mathtt{A}(a,s)$\tabularnewline
$\mathtt{B}_{\downarrow}(a,s)$ & $\mathtt{A}(a,s)$ & $\no$\tabularnewline
$\mathtt{B}_{\updownarrow}(a,s)$ & $\yes$ & $\no$\tabularnewline
\end{tabular}
\par\end{center}

\noindent The table also refers to the authentic agent type $\mathtt{A}$
to make it visually explicit in which cases the Up- and Downvote Boosters
behave authentically.

\subsubsection*{Distorters.}

Distorters seek to create noise among the votes by, on cue, voting
against their beliefs about quality. They vote in a coordinated, but
non-partisan fashion: triggered by $p_{3}$, they vote contrary to
their private beliefs about quality ($p_{1}$). As with $p_{2}$,
we think of $p_{3}$ as encoding a property of the post distinct from
quality, such as, e.g., tag, source or viewpoint. The $\mathtt{D}$
agents seek to water down the majority view and damper public impressions
of consensus, thus exhibiting one form of \emph{concern trolling }\citep{Goerzen2019}.

The main \emph{Upvote Distorter} $\mathtt{D}_{\uparrow}$ votes authentically
(to hide) except when they believe the post has $p_{3}$ but not $p_{1}$:
then they distort by voting contrary to their belief about $p_{1}$
(e.g., they upvote low quality posts of a given viewpoint to dampen
consensus impressions). For auxiliaries, the \emph{Downvote Distorter
}$\mathtt{D}_{\downarrow}$ `inverts' $\mathtt{D}_{\uparrow}$:
they vote authentically except when believing the post has both $p_{1}$
and $p_{3}$; then they distort by voting contrary to their beliefs
about quality. The\emph{ Both} \emph{Distorter} $\mathtt{D}_{\updownarrow}$
join the inauthentic behaviors of $\mathtt{D}_{\uparrow}$ and $\mathtt{D}_{\downarrow}$:
if they believe the post has $p_{3}$, then they vote contrary to
their $p_{1}$ beliefs (e.g., to always sow distrust about content
from a given source, or of a given viewpoint). The vote of an agent
$a$ of type $\mathtt{D}_{i\in\{\uparrow,\downarrow,\updownarrow\}}$
in state $s$ is $\mathtt{D}_{i}(a,s)$ given by
\begin{center}
\begin{tabular}{c|c|c|c}
 & $\genfrac{}{}{0pt}{1}{B_{a}(p_{3})=\yes,\text{ and}}{B_{a}(p_{1})=\yes\phantom{,\text{ and}}}$ & $\genfrac{}{}{0pt}{1}{B_{a}(p_{3})=\yes,\text{ and}}{B_{a}(p_{1})=\no\phantom{,\text{an}}}$ & $B_{a}(p_{3})=\no$\tabularnewline[4pt]
\hline 
$\mathtt{D}_{\uparrow}(a,s)$ & $B_{a}(p_{1})$ & $\no\cdot B_{a}(p_{1})$ & $\mathtt{A}(a,s)$\tabularnewline
$\mathtt{D}_{\downarrow}(a,s)$ & $\no\cdot B_{a}(p_{1})$ & $B_{a}(p_{1})$ & $\mathtt{A}(a,s)$\tabularnewline
$\mathtt{D}_{\updownarrow}(a,s)$ & $\no\cdot B_{a}(p_{1})$ & $\no\cdot B_{a}(p_{1})$ & $\mathtt{A}(a,s)$\tabularnewline
\end{tabular}
\par\end{center}

\subsubsection*{Lone Wolfs.}

Lone wolfs also create noise among the votes by voting against their
beliefs about quality. They do so exactly as the distorters, but without
coordination through $p_{3}$. We interpret these agents as individual
users that—cued by a personal property—upvote contra their beliefs
about quality ($\mathtt{L}_{\uparrow}$, main, \emph{Upvote Lone Wolf}),
e.g., out of sympathy, downvote contra their beliefs about quality
($\mathtt{L}_{\downarrow}$, aux., \emph{Downvote Lone Wolf}), e.g.,
out of anger or spite, or both ($\mathtt{L}_{\updownarrow}$, aux.,
\emph{Both Lone Wolf}).

Instead of voting given shared property $p_{3}$, a lone wolf, i.e.,
an agent $a$ of type $\mathtt{L}_{i\in\{\uparrow,\downarrow,\updownarrow\}}$,
votes on a personal property $p_{a}\in\{\no,\yes\}$, believing $B_{a}(p_{a})\in\{\no,\yes\}$
with $P(B_{a}(p_{a})=p_{a})=1$. For all $a,b\in A$, properties $p_{a},p_{b}$
are sampled as $p_{3}$, but if $a\neq b$, $p_{a}$ and $p_{b}$
are sampled independently. The voting rules for each $\mathtt{L}_{i}$,
$i\in\{\uparrow,\downarrow,\updownarrow\}$, is obtained by replacing
$p_{3}$ with $p_{a}$ in the table for $\mathtt{D}_{i}$.

\subsection{Majority Vote and Correctness}

We are interested in how agent populations' votes fair with respect
to \emph{majority correctness}, both before (baseline experiments)
and after we have applied our two jury selection procedures\emph{.}
A \emph{jury}\textbf{ }is a set of agents $J\subseteq A$. Let $(v_{a})_{a\in J}$,
$v_{a}\in\{\yes,\no\}$ be a \emph{voting profile} of $J$ with respect
to the post. The \emph{majority vote} of $(v_{a})_{a\in J}$ is whichever
of $\yes$ and $\no$ that gets more votes, tie-breaking to $\yes$,
giving the post the benefit of doubt. I.e., the majority vote of $(v_{a})_{a\in J}$
is $\no$ if $\sum_{a\in J}v_{a}<\zero$, else $\yes$. The majority
vote is \emph{correct} if it equals the post's quality, $p_{1}\in\{\yes,\no\}$.
Finally, the \emph{majority correctness score}\textbf{ }(\textsc{mcs})\textbf{
}of a jury over a set of voting rounds is the percentage of correct
majority votes of the jury in those rounds. The \textsc{mcs} of a
jury is a measure of its competence with respect to tracking quality,
and is the jury performance indicator of interest in this paper.

\subsection{\label{subsec:Parameters-and-Generated-Dataset}Parameters and Generated
Dataset}

Using R to implement the ABM,\footnote{We implemented the ABM from the ground up to retain freedom in agent
design and as the simplicity of the encoded behavior and generated
data do not invoke advanced features of existing ABM simulation packages
and programs, such as Netlogo \citep{Wilensky_Netlogo1999}, Laputa
\citep{angere2010knowledge,olsson2013bayesian}, or Hashkat \citep{ryczko2017hashkat}.} we chose three \emph{noise level} parameter combinations for the
sampling of properties:
\begin{center}
\begin{tabular}{c|c|c|c}
 & $P(p_{1}=\yes)$ & $P(p_{2}=\yes)$ & $P(p_{3}=\yes)$\tabularnewline[4pt]
\hline 
\textsc{low} & $0.75$ & $0.75$ & $0.9$\tabularnewline
\textsc{mid} & $0.75$ & $0.5$ & $0.5$\tabularnewline
\textsc{high} & $0.75$ & $0.1$ & $0.1$\tabularnewline
\end{tabular}
\par\end{center}

\noindent These noise levels were chosen to produce different voting
patterns, and to introduce varying degrees of coordination and correlations
among votes, in turn producing three levels of difficulty for vote-based
agent classification. Quality ($p_{1}$) is fixed across levels, leaving
authentic agents unaffected. Inauthentic agents perform less (coordinated)
inauthentic activities in higher levels, as $p_{2}$ and $p_{3}$
decrease. They thus mimic authentic agents more (more noise), raising
the difficulty of classification. The sampling of $p_{2}$ is asymmetric
to avoid mirrored results in low and mid noise for $\mathtt{B}_{\downarrow}$
and $\mathtt{B}_{\uparrow}$ given that booster agents solely rely
on $p_{2}$. We chose a symmetric setup for $P(p_{3}=\yes)$ as distorters
and lone wolfs' votes are not solely determined by $p_{3}$, but influenced
by the sampling of $p_{1}$, too, hence making completely mirrored
votes less likely. For each noise level, we performed $100$ runs,
each based on a random seed and with voting rounds $r=1000$, producing
a dataset with $3\times100,000$ (state, vote profile) pairs. Each
was done for $\mathcal{P}_{\mathsf{\mathtt{Full}}}$, thus counting
$1900$ agents: $1000$ authentic and $100$ of each inauthentic type.
Sec.~\ref{subsec:Baseline-Majority-Correctness} displays diverse
population ratios that explore the effect of authentic agents in minority
and majority on \textsc{mcs}. Throughout, results are based on and
evaluated against a datasubset with $r=500$. As all runs and rounds
are independent, choosing fewer or more voting rounds is without problem.
Other values of $r$ are mentioned explicitly when robustness checks
are discussed.

\subsection{\label{subsec:Baseline-Majority-Correctness}Baseline Majority Correctness
Scores}

\begin{figure}
\resizebox{\columnwidth}{!}{\input{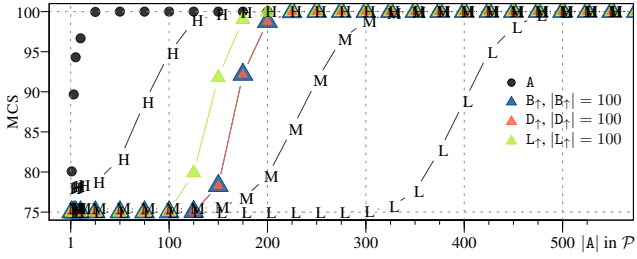}}

\caption{\label{fig:plot-A} Baseline\textsc{ mcs} of $\mathcal{P}$ with $|\mathtt{A}|=1,$3,5,10,25,50,$...$,
for $p_{2}=p_{3}=\yes$. Colored: populations with single inauthentic
type in low noise. \textsc{l}/\textsc{m}/\textsc{h}: populations with
multiple inauthentic types $\mathcal{P}=\{\mathtt{B}_{i},\mathtt{D}_{i},\mathtt{L}_{i}\}_{i\in\{\uparrow,\downarrow,\updownarrow\}}$,
with $|\mathtt{B}_{i}|=|\mathtt{D}_{i}|=|\mathtt{L}_{i}|=100$ in
low \textsc{(l}), mid \textsc{(m}), high \textsc{(h}) noise. Find
\textsc{mcs} of $\mathcal{P}_{\mathbf{\mathtt{All}}}$, $\mathcal{P_{\mathtt{B}_{\mathrm{\uparrow}}}}$,
$\mathcal{P_{\mathtt{D}_{\uparrow}}}$, $\mathcal{P}_{\mathtt{L_{\uparrow}}}$
at $|\mathtt{A}|=100$.}
\end{figure}

\begin{figure}
\resizebox{\columnwidth}{!}{\input{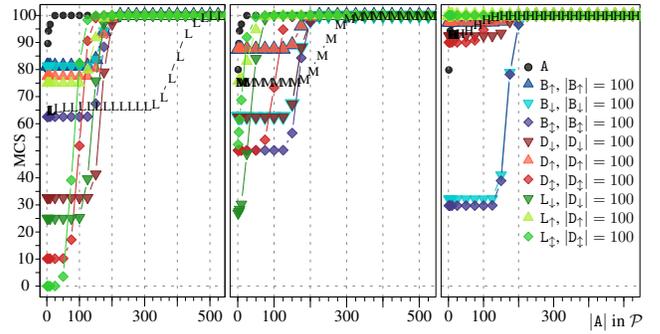}}

\caption{\label{fig:plot-B} Baseline\textsc{ mcs} of $\mathcal{P}$ with $|\mathtt{A}|=1,$3,5,10,25,50,$...$
for low (left), mid (mid), high (right) noise and all agent types.
\textsc{l}/\textsc{m}/\textsc{h}: $\mathcal{P}=\{\mathtt{B}_{i},\mathtt{D}_{i},\mathtt{L}_{i}\}_{i\in\{\uparrow,\downarrow,\updownarrow\}}$,
$|\mathtt{B}_{i}|=|\mathtt{D}_{i}|=|\mathtt{L}_{i}|=100$ in low \textsc{(l}),
mid \textsc{(m}), high \textsc{(h}) noise. Find \textsc{mcs} of $\mathcal{P}_{\mathbf{\mathtt{All}}}$,
$\mathcal{P_{\mathtt{B}_{\mathrm{\uparrow}}}}$, $\mathcal{P_{\mathtt{D}_{\uparrow}}}$,
$\mathcal{P}_{\mathtt{L_{\uparrow}}}$ at $|\mathtt{A}|=100$.}
\end{figure}

To showcase varying populations' behaviors, we illustrate two sets
of baseline \textsc{mcs} results in Figures \ref{fig:plot-A} and
\ref{fig:plot-B}.

Figure~\ref{fig:plot-A} shows $7$ populations' \textsc{mcs}s as
a function of the number $|\mathtt{A}|$ of authentic agents in the
population. As expected from the Condorcet Jury Theorem, the \textsc{mcs}
of authentic agents alone converges to 100\%, with 25 agents sufficing.
This is representative for all noise levels, as noise does not affect
authentic agents. Figure~\ref{fig:plot-A} is filtered to rounds
with $p_{2}=p_{3}=\yes$, so $\mathtt{B}_{\uparrow}$ and $\mathtt{D}_{\uparrow}$
are `actively inauthentic' (and both always upvote). Given this
filter, the figure is representative for all noise levels for $\mathtt{B}_{\uparrow}$
and $\mathtt{D}_{\uparrow}$. The effect of $\mathtt{L}_{\uparrow}$
is level specific (but unaffected by the filter).

We make three observations concerning the main, upvoting inauthentic
agents $\mathtt{B}_{\uparrow},\mathtt{D}_{\uparrow}$ and $\mathtt{L}_{\uparrow}$
of Figure~\ref{fig:plot-A}. \textbf{First}, the left-most part of
Figure~\ref{fig:plot-A} shows populations with $|\mathtt{A}|=1$,
a very hospitable environment for inauthentic activity. Here, each
of $\mathtt{B}_{\uparrow},\mathtt{D}_{\uparrow}$ and $\mathtt{L}_{\uparrow}$
exhibit a \textsc{mcs} of $75\%$. This is an artifact of how their
behavior interacts with the sampling frequency for $p_{1}$. For $\mathtt{B}_{\uparrow}$
and $\mathtt{D}_{\uparrow}$, the \textsc{mcs} of $75\%$ follows
as Figure~\ref{fig:plot-A} is filtered for $p_{2}=p_{3}=1$, and
only contains rounds where both always upvote. As $p_{1}=75\%$, they
are thus correct $75\%$ of the time. Though not all $\mathtt{L}_{\uparrow}$
always upvote in these rounds, they do so individually with a $96.5\%$
chance (assuming average $C_{a}(p_{1})=0.8$). As a group, they thus
sway the majority vote to $\yes$ with high probability, again correct
with $75\%$. \textbf{Second}, $\mathtt{B}_{\uparrow},\mathtt{D}_{\uparrow}$
and $\mathtt{L}_{\uparrow}$ each exhibit their maximal lowering effect
on the \textsc{mcs} while $|\mathtt{A}|=100$. This is a motivating
factor in focusing on populations with $|\mathtt{A}|=100$, the \textsc{mcs}s
of which we return to in Table~\ref{tab:MCS:E+F_results}. \textbf{Third},
for $|\mathtt{A}|>100$, $\mathtt{B}_{\uparrow}$ and $\mathtt{D}_{\uparrow}$
negatively influence the \textsc{mcs} identically, as both upvote
in the shown rounds, with their effect declining from a \textsc{mcs}
of $75\%$ at $|\mathtt{A}|=125$ to an \textsc{mcs} of $100\%$ by
$|\mathtt{A}|=225$. For $|\mathtt{A}|>100$, to form an incorrect
majority, inauthentic agents must be `aided' by authentic agents
that happen to vote incorrectly. The probability that enough such
exist to overcome the correctly voting authentic agents drops as $|\mathtt{A}|$
grows. With $|\mathtt{A}|\geq225$, $\mathtt{B}_{\uparrow}$ and $\mathtt{D}_{\uparrow}$
are seen to have lost all effect. $\mathtt{L}_{\uparrow}$ have a
less robust effect, as they vote in an uncoordinated fashion, and
are thus more quickly outnumbered by authentic agents' votes.

Finally, the effect of the $900$ inauthentic agents jointly drops
with higher noise levels, i.e., with decreased activity. In the high
activity case (low noise), the 900 inauthentic agents seem `overwhelmed'
already by between 325 and 475 authentic agents. This is correct on
the aggregate level, but $900$ inauthentic agents do not equate $900$
inauthentic actions: given the filter, some types act authentically
always ($\mathtt{B}_{\downarrow}$) or sometimes ($\mathtt{D}_{\uparrow},\mathtt{D}_{\downarrow},\mathtt{L}_{\uparrow},\mathtt{L}_{\downarrow},\mathtt{L}_{\updownarrow}$).
Additionally, some types partially cancel each other (e.g., $\mathtt{D}_{\uparrow}$
and $\mathtt{L}_{\downarrow}$) or even themselves (e.g., $\mathtt{D}_{\updownarrow}$)
out.

Figure~\ref{fig:plot-B} shows \textsc{mcs} summary plots of all
inauthentic agent types in isolation and jointly, as a function of
$|\mathtt{A}|$, not filtered for properties. As noise increases,
the figure evinces how inauthentic agents' impact on \textsc{mcs}
decreases. Note how in low noise, agent types $\mathtt{D}_{\updownarrow}$,
$\mathtt{D}_{\uparrow}$, $\mathtt{L}_{\updownarrow}$, and $\mathtt{L}_{\uparrow}$
are more effective than $\mathtt{B}_{i}$ for each $i\in\{\uparrow,\downarrow,\updownarrow\}$
in lowering the \textsc{mcs} as the former agent types directly counteract
correct majority voting concerning quality ($p_{1}$). The picture
flips in the high noise level given how $p_{1}$, $p_{2}$, and $p_{3}$
are sampled (Sec.~\ref{subsec:Parameters-and-Generated-Dataset}).

\section{\label{sec:Classification}Classification}

Our jury selection procedures \textsc{(gmm} and \textsc{km} \textsc{jsp}s)
classify the set of agents into two agent groups: authentic and inauthentic.\textbf{
}Each jury selection procedure invokes a classifier method that decomposes
the ABM data into singular values (SVD), applies a clustering strategy—either
a Gaussian Mixture Model (\textsc{gmm}) or the $k$-means algorithm
(\textsc{km})—and labels agents using logistic regression on the quality
property $p_{1}$ of the post voted on.

We assume $p_{1}$ known, as we know of the general setting: the agents
vote on quality. We do not assume knowledge of $p_{2}$ and $p_{3}$,
or even of their existence. The input dataset consists of binary votes
of $n$ agents over a given number of voting rounds $r$, where $r>n$
is not a requirement regarding the machinery. Yet the more observations
$r$, the better we cluster. Data requirements are thus feasible,
in contrast to the $\chi^{2}$ test suggested by \citet{GaleazziRendsvigSlavkovik2019}
that requires $p_{1}$ known plus at least $1$ observation for each
of the $2^{n}$ possible voting round outcomes.

\begin{figure}[t]
\resizebox{\columnwidth}{!}{\input{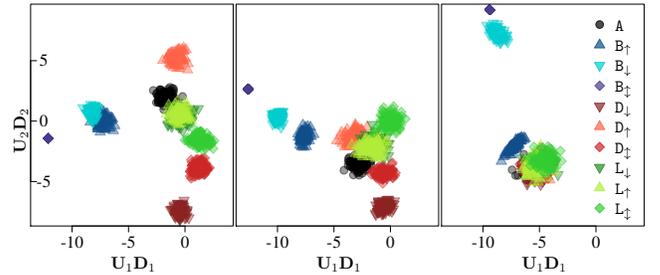}}\caption{\label{fig:figure_Scatterplot} Exemplary, representative scatterplot
of $\mathbf{U}_{q}\mathbf{D}_{q}$ for $n=1000$ for $\mathcal{P}_{\mathtt{All}}$
and $r=500$, in low (left), mid (mid), high (right) noise.}
\end{figure}

For each ABM run, the classification analysis is performed on five
resampled (with replacement) datasets with $r=500$ and $n$ either
$1000$ for $\mathcal{P}_{\mathtt{All}}$ or $200$ for $\mathcal{P}_{\mathtt{B}_{\uparrow}}$,
$\mathcal{P}_{\mathtt{D}_{\uparrow}}$, and $\mathcal{P}_{\mathtt{L}_{\uparrow}}$.
For each of the bootstrapped datasets, we calculate the Singular Value
Decomposition (SVD) $\mathbf{X=UDV}^{T}$ of the $n\times n$ sample
correlation matrix $\mathbf{X}$ of the vote data. For clustering,
we consider the first $q=2$ dimensions' eigenvectors, i.e., the first
two columns of the $n\times p$ orthogonal matrix $\mathbf{U}$ where
$n=p$, weighted with the corresponding eigenvalue collected in the
diagonal $p\times p$ matrix $\mathbf{D}$. Hence, we cluster on the
$q$ partial components $\mathbf{U}_{q}\mathbf{D}_{q}$ \citep{hastie_09_elements-of.statistical-learning}.
Figure~\ref{fig:figure_Scatterplot}
 shows the scatterplots of $\mathbf{U}_{q}\mathbf{D}_{q}$, illustrating
more blurred clustering environments as the noise level increases
from low to high.

We contrast the probabilistic Gaussian Mixture Model (\textsc{gmm})
and the deterministic $k$-means (\textsc{km}) algorithm for clustering
the components. The soft clustering \textsc{gmm} is more memory-intensive,
while the hard clustering \textsc{km} algorithm is faster. We choose
\textsc{gmm} and $k$-means as they are among the simplest, most well-known,
and most efficiently implementable unsupervised clustering methods
\citep{hastie_09_elements-of.statistical-learning}. Both cluster
the weighted eigenvectors into $k$ groups, $k=2,...,20$ (in testing,
20 proved sufficient as upper bound). In the \textsc{gmm}, $k$ is
chosen by maximizing the $\log$-likelihood according to the Bayesian
Information Criterion (BIC). The BIC penalizes the number of parameters
more heavily than Akaikes Information Criterion, aiming for a model
fit with fewer parameters to avoid overfitting \citep{mclust}. In
the \textsc{km} algorithm, $k$ is estimated with the gap statistic,
which compares the change in the within-cluster dispersion with that
under a reference null distribution \citep{Tibshirani2001}.

Having clustered the data into $k$ groups, the mean vote per voting
round of those agents clustered together—i.e., the row sums of $k$
subsets of the vote data, viz. $k$ $r\times1$ vectors—are used in
a logistic regression model with the two-level factor $p_{1}$ as
the response variable. Put differently, the $k$ coefficients refer
to the clusters' mean vote per voting round given the quality of posts.
To select those clusters comprising inauthentic agents, we add the
lasso penalty term to the optimization, $\sum_{j=1}^{k}||\beta||_{j}$
with $k$ predictor variables (clusters), as implemented in the R
package \texttt{glmnet} \citep{glmnet}. Coefficients consequently
shrunk to $0$ when regressing on $p_{1}$ receive the label `inauthentic'.
Coefficients \emph{not} shrunk to $0$ receive the label `authentic'.
Lasso regularization was chosen over the ridge regularization as the
former shrinks coefficients to $0$ and thereby imposes sparseness.
In contrast, the ridge penalty never fully removes variables. Coefficients
shrunk to $0$ accordingly do not play an important role when regressing
on $p_{1}$ and therefore receive the label `inauthentic'. These
labels are then forwarded to the agents found in each cluster.

Note that logistic regression is applied in a non-standard way. In
this paper, the goal of the logistic regression is \emph{not} to predict
each vote per voting round into the categorical dependent variable
$p_{1}$, in contrast to classical approaches where the dependent
variable describes the classes in which one is interested. We seek
to classify each agent as inauthentic or authentic which we do via
hard classification through shrinking components to $0$ and an additional
labeling step. This makes traditional classifier metrics like a receiver
operating characteristic curve (ROC curve) and a corresponding area
under the curve score (AUC score) inapplicable. Instead, Sec.~\ref{subsec:Classification-Results}
discusses false positive and false negative classification errors
to transparently and separately assess the two desiderata vox populi
and precaution.

Once the classification analysis is completed on all $5$ bootstrapped
datasets, each agent has been classified as either `authentic' or
`inauthentic' $5$ times. Only if an agent received the `authentic'
label at least $4$ out of $5$ times, the overall `authentic' label
will be granted. Else, the agent is overall classified as `inauthentic'.
The $\nicefrac{4}{5}$ classification threshold was fixed pragmatically
to balance runtime efficiency and precaution\emph{ }against inauthentic
influence. Simple majority would exhibit less precaution and more
vox populi, while a $\nicefrac{19}{20}$ classification threshold
(95\%) would heavily increase runtime. We discuss this modeling choice
further in the final remarks (Sec.~\ref{subsec:Assumptions}).

\subsection{\label{subsec:Classification-Results}Classification Results}

In order to evaluate the \textsc{gmm} and \textsc{km} classifier methods,
we first inspect the misclassification results for $\mathcal{P_{\mathtt{All}}}$
for $r=500$, second comment on selected robustness observations,
and third examine classifier accuracy for smaller sub-populations
$\mathcal{P_{\mathtt{B}_{\mathrm{\uparrow}}}}$, $\mathcal{P_{\mathtt{D}_{\uparrow}}}$,
and $\mathcal{P_{\mathtt{L}_{\uparrow}}}$ for $r=500$.

\begin{table*}
\setlength\tabcolsep{4pt}
\noindent \begin{centering}
\begin{tabular}{clcccccccccccc}
 &  & \multicolumn{4}{c}{\textsc{low}} & \multicolumn{4}{c}{\textsc{mid}} & \multicolumn{4}{c}{\textsc{high}}\tabularnewline
\cmidrule{3-14} \cmidrule{4-14} \cmidrule{5-14} \cmidrule{6-14} \cmidrule{7-14} \cmidrule{8-14} \cmidrule{9-14} \cmidrule{10-14} \cmidrule{11-14} \cmidrule{12-14} \cmidrule{13-14} \cmidrule{14-14} 
 &  & $\,\mathtt{All}$ & $\mathtt{B}_{\uparrow}$ & $\mathtt{D}_{\uparrow}$ & $\mathtt{L}_{\uparrow}$ & $\,\mathtt{All}$ & $\mathtt{B}_{\uparrow}$ & $\mathtt{D}_{\uparrow}$ & $\mathtt{L}_{\uparrow}$ & $\,\mathtt{All}$ & $\mathtt{B}_{\uparrow}$ & $\mathtt{D}_{\uparrow}$ & $\mathtt{L}_{\uparrow}$\tabularnewline
\multirow{2}{*}{\textsc{gmm}} & {\footnotesize{}Authentic} & $\tfrac{.04}{(.04)}$ & $-$ & $-$ & $\tfrac{.03}{(.05)}$ & $\tfrac{.11}{(.13)}$ & $-$ & $\tfrac{.00}{(.008)}$ & $\tfrac{.04}{(.03)}$ & $\tfrac{.5}{(.26)}$ & $\tfrac{.01}{(.02)}$ & $\tfrac{.01}{(.02)}$ & $\tfrac{.09}{(.06)}$\tabularnewline
 & {\footnotesize{}Inauthentic} & $\tfrac{.03}{(.03)}$ & $\tfrac{.13}{(.00)}$ & $-$ & $-$ & $\tfrac{.04}{(.04)}$ & $\tfrac{.13}{(.00)}$ & $-$ & $\tfrac{.03}{(.12)}$ & $\tfrac{.35}{.17}$ & $\tfrac{.13}{(.002)}$ & $-$ & $\tfrac{.81}{(.11)}$\tabularnewline\addlinespace[2pt]
\midrule
\addlinespace[2pt]
\multirow{2}{*}{\textsc{km}} & {\footnotesize{}Authentic} & $\tfrac{.2}{(.4)}$ & $-$ & $-$ & $\tfrac{.002}{(.006)}$ & $\tfrac{.07}{(.17)}$ & $\tfrac{.01}{(.03)}$ & $\tfrac{.00}{(.01)}$ & $\tfrac{.016}{(.016)}$ & $\tfrac{.31}{(.29)}$ & $\tfrac{.03}{(.004)}$ & $\tfrac{.05}{(.07)}$ & $\tfrac{.01}{(.03)}$\tabularnewline
 & {\footnotesize{}Inauthentic} & $\tfrac{.43}{(.28)}$ & $\tfrac{.13}{(.00)}$ & $\tfrac{.05}{(.06)}$ & $\tfrac{.00}{(.001)}$ & $\tfrac{.48}{(.12)}$ & $\tfrac{.13}{(.003)}$ & $\tfrac{.05}{(.06)}$ & $\tfrac{.06}{(.11)}$ & $\tfrac{.56}{(.16)}$ & $\tfrac{.13}{(.01)}$ & $\tfrac{.44}{(.41)}$ & $\tfrac{.97}{(.07)}$\tabularnewline
\end{tabular}
\par\end{centering}
\noindent \setlength\tabcolsep{6pt}\bigskip{}
\smallskip{}

\caption{\label{tableD}\textsc{gmm} and \textsc{km} mean misclassification
and standard deviation ($\tfrac{\text{Mean}}{(\text{SD})}$; `$-$'
is short for $\tfrac{.00}{(.00)}$) of authentic (\textsc{g}) and
inauthentic (\textsc{n}) agents in populations $\mathcal{P}_{\mathtt{All}}$,
$\mathcal{P}_{\mathtt{B}_{\uparrow}}$, $\mathcal{P}_{\mathtt{D}_{\uparrow}}$
and $\mathcal{P}_{\mathtt{L}_{\uparrow}}$, for each noise level.
E.g., for low noise in $\mathcal{P}_{\mathtt{D}_{\uparrow}}$, \textsc{gmm}
perfectly classifies, while \textsc{km} misclassifies no authentic
agents, but $5\%$ inauthentic agents (SD = $6\%$).}
\end{table*}

\begin{figure}[t]
\centering{}\resizebox{\columnwidth}{!}{\input{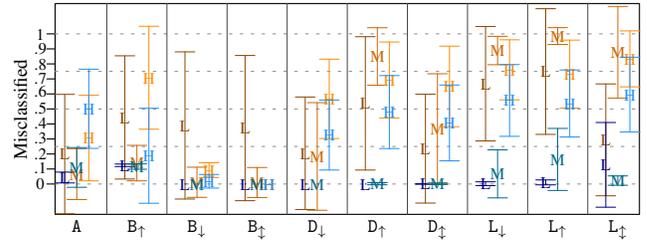}}\caption{\label{fig:figureC} \textsc{gmm} (blue) and \textsc{km }(orange)
mean misclassification and standard deviation in $\mathcal{P}_{\mathtt{All}}$
in low (\textsc{l}), mid (\textsc{m}), and high (\textsc{h}) noise.}
\end{figure}

First, in population $\mathcal{P_{\mathtt{All}}}$, \textsc{gmm} classifies
well in the low (mid) noise case, accurately misclassifying only $4$\%
($11$\%) of authentic agents as inauthentic, and $3$\% ($4$\%)
of inauthentic agents as authentic (Table~\ref{tableD}), exhibiting
both vox populi and precaution. As expected, classifier accuracy reduces
in the high noise case given that inauthentic agents hide and mimic
authentic behavior, i.e., often vote authentically and are accordingly
difficult to detect. However, here the inauthentic agents' impact
on majority correctness scores is limited (Figure~\ref{fig:plot-B})
despite a $35$\% false negative error. Indeed, as Figure~\ref{fig:plot-B}
suggests, it is $\mathtt{B}_{\downarrow}$ and $\mathtt{B}_{\updownarrow}$
agents that negatively affect the \textsc{mcs} to the largest extent
in the high noise case, and both \textsc{gmm} and \textsc{km} identify
these agent groups accurately as inauthentic (Figure~\ref{fig:figureC}).
Moreover, classification results in Figure~\ref{fig:figureC} and
Table~\ref{tableD} show how \textsc{gmm} outperforms \textsc{km}
in all noise levels with regard to identifying inauthentic agents,
exhibiting less false negative misclassification. Thus, \textsc{gmm}
overall clusters more precautiously than \textsc{km}.

Second, robustness checks given $\mathcal{P_{\mathtt{All}}}$ show
differences between \textsc{gmm} and \textsc{km}. Results based on
fewer observations ($r=250$ instead of $r=500)$ affect false negative
errors less for \textsc{gmm}, but notably for \textsc{km.} E.g., \textsc{gmm}
still does not misclassify any booster or distorter agents, while
\textsc{km'}s false negative errors ($\tfrac{\text{Mean}}{(\text{SD})}$)
rise in low noise as follows
\begin{center}
\setlength\tabcolsep{1.1pt}\resizebox{\columnwidth}{!}{
\begin{tabular}{ccc|cc|cc|cc|cc|cc}
 & \multicolumn{2}{c|}{$\mathtt{B}_{\uparrow}$} & \multicolumn{2}{c|}{$\mathtt{B}_{\downarrow}$} & \multicolumn{2}{c|}{$\mathtt{B}_{\updownarrow}$} & \multicolumn{2}{c|}{$\mathtt{D}_{\uparrow}$} & \multicolumn{2}{c|}{$\mathtt{D}_{\downarrow}$} & \multicolumn{2}{c}{$\mathtt{D}_{\updownarrow}$}\tabularnewline
\cline{2-13} \cline{3-13} \cline{4-13} \cline{5-13} \cline{6-13} \cline{7-13} \cline{8-13} \cline{9-13} \cline{10-13} \cline{11-13} \cline{12-13} \cline{13-13} 
{\footnotesize{}$r$} & {\scriptsize{}$500$} & {\scriptsize{}$250$} & {\scriptsize{}$500$} & {\scriptsize{}$250$} & {\scriptsize{}$500$} & {\scriptsize{}$250$} & {\scriptsize{}$500$} & {\scriptsize{}$250$} & {\scriptsize{}$500$} & {\scriptsize{}$250$} & {\scriptsize{}$500$} & {\scriptsize{}$250$}\tabularnewline
\textsc{km} & $\tfrac{.44}{(.41)}$ & $\tfrac{.56}{(.43)}$ & $\tfrac{.39}{(.5)}$ & $\tfrac{.52}{(.5)}$ & $\tfrac{.37}{(.48)}$ & $\tfrac{.51}{(.5)}$ & $\tfrac{.54}{(.44)}$ & $\tfrac{.68}{(.4)}$ & $\tfrac{.39}{(.5)}$ & $\tfrac{.52}{(.5)}$ & $\tfrac{.24}{(.36)}$ & $\tfrac{.34}{(.41)}$\tabularnewline
\end{tabular}}\vspace*{.1cm}
\par\end{center}

\noindent with similar trends observable in mid and high noise. Based
on more observations ($r=750$ ($r=1000$) instead of $r=500$), both
\textsc{gmm }and\textsc{ km }classify all inauthentic agent types
but $\mathtt{L}_{\updownarrow}$ more precautiously in difficult high
noise environments. Misclassification in high noise for the main inauthentic
agent types improve thusly:
\begin{center}
\setlength\tabcolsep{3pt}\resizebox{\columnwidth}{!}{
\begin{tabular}{cccc|ccc|ccc}
 & \multicolumn{3}{c|}{$\mathtt{B}_{\uparrow}$} & \multicolumn{3}{c|}{$\mathtt{D}_{\uparrow}$} & \multicolumn{3}{c}{$\mathtt{L}_{\uparrow}$}\tabularnewline
\cline{2-10} \cline{3-10} \cline{4-10} \cline{5-10} \cline{6-10} \cline{7-10} \cline{8-10} \cline{9-10} \cline{10-10} 
{\footnotesize{}$r$} & {\scriptsize{}$500$} & {\scriptsize{}$750$} & {\scriptsize{}$1000$} & {\scriptsize{}$500$} & {\scriptsize{}$750$} & {\scriptsize{}$1000$} & {\scriptsize{}$500$} & {\scriptsize{}$750$} & {\scriptsize{}$1000$}\tabularnewline
\textsc{gmm} & $\tfrac{.19}{(.32)}$ & $\tfrac{.09}{(.2)}$ & $\tfrac{.11}{(.24)}$ & $\tfrac{.47}{(.24)}$ & $\tfrac{.29}{(.19)}$ & $\tfrac{.25}{(.16)}$ & $\tfrac{.54}{(.22)}$ & $\tfrac{.36}{(.2)}$ & $\tfrac{.33}{(.17)}$\tabularnewline
\textsc{km} & $\tfrac{.7}{(.34)}$ & $\tfrac{.5}{(.39)}$ & $\tfrac{,49}{(.38)}$ & $\tfrac{.69}{(.25)}$ & $\tfrac{.46}{(.26)}$ & $\tfrac{.43}{(.28)}$ & $\tfrac{.73}{(.23)}$ & $\tfrac{.53}{(.24)}$ & $\tfrac{.51}{(.25)}$\tabularnewline[2pt]
\end{tabular}}
\par\end{center}

\noindent However, in the same case, false positive errors increase:
authentic agent misclassification worsens from $\tfrac{.5}{(.26)}$
to $\tfrac{.67}{(.24)}$ ($\tfrac{.73}{(.19)}$) for \textsc{gmm},
and from $\tfrac{.3}{(.29)}$ to $\tfrac{.48}{(.3)}$ ($\tfrac{.54}{(.32)}$)
for \textsc{km}. In contrast to \textsc{gmm}, \textsc{km} demonstrates
robustness shortcomings as its classification accuracy notably improves
with increased $|\mathtt{A}|=1000$. This difference is pronounced
in false positive errors in low (mid) noise levels for $r=500$: \textsc{km}
improves from mean misclassification $\tfrac{.2}{(.4)}$ to $\tfrac{.00}{(.00)}$
($\tfrac{.07}{(.17)}$ to $\tfrac{.01}{(.01)}$), while \textsc{gmm}
misclassification changes from $\tfrac{.04}{(.04)}$ to $\tfrac{.1}{(.06)}$
($\tfrac{.11}{(.13)}$ to $\tfrac{.1}{(.07)}$). Neither notably improves
false positive errors in the high noise case, but the misclassification
exhibits lower SD (\textsc{km}: $\tfrac{.3}{(.29)}$ to $\tfrac{.31}{(.15)}$,
\textsc{gmm}: $\tfrac{.5}{(.26)}$ to $\tfrac{.47}{(.16)}$). Full
robustness results can be produced with the Supplementary Material.

Third, in small sub-populations $\mathcal{P_{\mathtt{B}_{\mathrm{\uparrow}}}}$
and $\mathcal{P_{\mathtt{D}_{\uparrow}}}$, we accurately classify
inauthentic agents as such without significant costs of false positive
errors: \textsc{gmm} weakly outperforms \textsc{km} throughout (it
does at least as good everywhere, and sometimes strictly better),
and strictly outperforms \textsc{km} in the high noise case (it does
strictly better everywhere), cf. Table~\ref{tableD}. In $\mathcal{P}_{\mathtt{L_{\uparrow}}}$,
both perform well in low and mid noise cases, however, fail to accurately
distinguish between $\mathtt{A}$ and $\mathtt{L}_{\uparrow}$ in
the high noise case, causing large false negative errors. Yet again,
$\mathtt{L}_{\uparrow}$ agents do not have a robust effect in watering
down \textsc{mcs}s (Figure~\ref{fig:plot-A}), given their uncoordinated
and camouflaging behavior.

\section{\label{sec:Jury-Selection-and}Jury Selection and Majority Correctness}

\begin{table*}
\setlength\tabcolsep{2pt}
\begin{centering}
\begin{tabular}{ccccccccccccccccc}
 & \multicolumn{3}{l}{\textsc{\enskip{}Filter: none}} & \multicolumn{3}{l}{\textsc{\enskip{}Filter: }${\scriptstyle p_{2}=p_{3}=\yes}$} & \multicolumn{5}{l}{\textsc{\enskip{}Filter: none}} & \multicolumn{5}{l}{\textsc{\enskip{}Filter: }${\scriptstyle p_{2}=p_{3}=\yes}$}\tabularnewline
 & $\mathtt{B}_{\uparrow}$ \colagt & \textsc{gmm} & \textsc{km} & $\mathtt{B}_{\uparrow}$ \colagt & \textsc{gmm} & \textsc{km} & $\mathtt{D}_{\uparrow}$ \colagt & \multicolumn{1}{c}{\textsc{gmm}} & \multicolumn{3}{c}{\textsc{km}} & $\mathtt{D}_{\uparrow}$ \colagt & \multicolumn{1}{c}{\textsc{gmm}} & \multicolumn{3}{c}{\textsc{km}}\tabularnewline
 & \textsc{b}{\footnotesize{}ase} & \textsc{b\,a\,w} & \textsc{b\,a\,w} & \textsc{b}{\footnotesize{}ase} & \textsc{b\,a\,w} & \textsc{b\,a\,w} & \textsc{b}{\footnotesize{}ase} & \textsc{b\,a\,w} & \textsc{b} & \textsc{a} & \textsc{w} & \textsc{b}{\footnotesize{}ase} & \textsc{b\,a\,w} & \textsc{b} & \textsc{a} & \textsc{w}\tabularnewline
\textsc{low} & $\tfrac{81.11}{(1.99)}$ & ${\scriptstyle -\,-\,-}$ & ${\scriptstyle -\,-\,-}$ & $\tfrac{74.83}{(2.75)}$ & ${\scriptstyle -\,-\,-}$ & ${\scriptstyle -\,-\,-}$ & $\tfrac{77.43}{(2.18)}$ & ${\scriptstyle -\,-\,-}$ & $-$ & $-$ & $-$ & $\tfrac{74.83}{(2.75)}$ & ${\scriptstyle -\,-\,-}$ & $-$ & $-$ & $-$\tabularnewline
\textsc{mid} & $\tfrac{87.50}{(1.39)}$ & ${\scriptstyle -\,-\,-}$ & ${\scriptstyle -\,-\,-}$ & $\tfrac{75.41}{(3.8)}$ & ${\scriptstyle -\,-\,-}$ & ${\scriptstyle -\,-\,-}$ & $\tfrac{87.62}{(1.6)}$ & ${\scriptstyle -\,-\,-}$ & $-$ & $-$ & $-$ & $\tfrac{75.41}{(3.8)}$ & ${\scriptstyle -\,-\,-}$ & $-$ & $-$ & $-$\tabularnewline
\textsc{high} & $\tfrac{97.58}{(.73)}$ & ${\scriptstyle -\,-\,-}$ & ${\scriptstyle -\,-\,-}$ & $\tfrac{77.17}{(21.54)}$ & ${\scriptstyle -\,-\,-}$ & ${\scriptstyle -\,-\,-}$ & $\tfrac{97.59}{(.73)}$ & ${\scriptstyle -\,-\,-}$ & $-$ & $\tfrac{99.89}{(.14)}$ & $\tfrac{97.59}{(.73)}$ & $\tfrac{77.17}{(21.54)}$ & ${\scriptstyle -\,-\,-}$ & $-$ & $\tfrac{99.20}{(3.52)}$ & $\tfrac{77.17}{(21.54)}$\tabularnewline
\midrule
 & $\mathtt{L}_{\uparrow}$ \colagt & \textsc{gmm} & \textsc{km} & $\mathtt{L}_{\uparrow}$ \colagt & \textsc{gmm} & \textsc{km} & \emph{$\mathtt{All}$}\colagt & \multicolumn{1}{c}{\textsc{gmm}} & \multicolumn{3}{c}{\textsc{km}} & \emph{$\mathtt{All}$}\colagt & \multicolumn{1}{c}{\textsc{gmm}} & \multicolumn{3}{c}{\textsc{km}}\tabularnewline
 & \textsc{b}{\footnotesize{}ase} & \textsc{b\,a\,w} & \textsc{b\,a\,w} & \textsc{b}{\footnotesize{}ase} & \textsc{b\,a\,w} & \textsc{b\,a\,w} & \textsc{b}{\footnotesize{}ase} & \textsc{b\,\,\,a\,\,\,w} & \textsc{b} & \textsc{a} & \textsc{w} & \textsc{b}{\footnotesize{}ase} & \textsc{b\,\,\,a\,\,\,w} & \textsc{b} & \textsc{a} & \textsc{w}\tabularnewline
\textsc{low} & $\tfrac{74.94}{(2.25)}$ & ${\scriptstyle -\,-\,-}$ & ${\scriptstyle -\,-\,-}$ & $\tfrac{74.93}{(2.77)}$ & ${\scriptstyle -\,-\,-}$ & ${\scriptstyle -\,-\,-}$ & $\tfrac{66.21}{(2.18)}$ & ${\scriptstyle -\,-}\tfrac{92.41}{(1.27)}$ & $-$ & $\tfrac{82.19}{(1.98)}$ & $\tfrac{64.21}{(2.27)}$ & $\tfrac{74.83}{(2.75)}$ & ${\scriptstyle -\,-}\tfrac{89.84}{(1.81)}$ & $-$ & $\tfrac{74.83}{(2.75)}$ & $\tfrac{74.83}{(2.75)}$\tabularnewline
\textsc{mid} & $\tfrac{99.98}{(.07)}$ & ${\scriptstyle -\,-\,-}$ & ${\scriptstyle -\,-\,-}$ & $\tfrac{99.99}{(.07)}$ & ${\scriptstyle -\,-\,-}$ & ${\scriptstyle -\,-\,-}$ & $\tfrac{75.12}{(2.08)}$ & ${\scriptstyle -\,-\,\,\,\,\,-}$\textsc{\,\,$\phantom{.}$} & $\tfrac{100.00}{(.02)}$ & $\tfrac{97.07}{(.77)}$ & $\tfrac{92.99}{(1.1)}$ & $\tfrac{75.41}{(3.8)}$ & ${\scriptstyle -\,-\,\,\,\,\,\,-}$\textsc{\,\,\,$\phantom{.}$} & $\tfrac{99.99}{(.07)}$ & $\tfrac{89.54}{(2.8)}$ & $\tfrac{78.43}{(3.72)}$\tabularnewline
\textsc{high} & $-$ & ${\scriptstyle -\,-\,-}$ & ${\scriptstyle -\,-\,-}$ & $-$ & ${\scriptstyle -\,-\,-}$ & ${\scriptstyle -\,-\,-}$ & $\tfrac{98.37}{(.45)}$ & ${\scriptstyle -\,-}\tfrac{99.90}{(.14)}$ & $\!-$ & $-$ & $\tfrac{99.95}{(.1)}$ & $\tfrac{94.18}{(11.34)}$ & ${\scriptstyle -\,-}\tfrac{91.02}{(12.79)}$ & $-$ & $-$ & $\tfrac{94.92}{(10.74)}$\tabularnewline
\end{tabular}
\par\end{centering}
\setlength\tabcolsep{6pt} 

\bigskip{}
\smallskip{}

\caption{\label{tab:MCS:E+F_results}\textbf{ }Mean \textsc{mcs} and standard
deviation ($\tfrac{\text{Mean}}{(\text{SD})}$; `$-$' is short
for $\tfrac{100.00}{(0.00)}$) for populations $\mathcal{P}_{\mathtt{B}_{\uparrow}},\mathcal{P}_{\mathtt{D}_{\uparrow}},\mathcal{P}_{\mathtt{L}_{\uparrow}}$
and $\mathcal{P}_{\mathtt{All}}$ in their baseline form (\textsc{b}{\footnotesize{}ase})
and in the best (\textsc{b}), average (\textsc{a}) and worst (\textsc{w})
cases for each of the \textsc{gmm} and \textsc{km} \textsc{jsp}s,
for $r=500$ either unfiltered or filtered so $\mathtt{B}_{\uparrow}$
and $\mathtt{D}_{\uparrow}$ are active ($p_{2}=p_{3}=\yes$). Each
of the 8 sub-tables (with tinted upper left corner for population
subscript) allows $i$) comparisons of a population's \textsc{mcs}
with those of the \textsc{jsp}s' best, average and worst case juries,
and $ii$) comparisons of the \textsc{mcs}s of the two \textsc{km}
and \textsc{gmm} \textsc{jsp}s.}
\end{table*}
The \textsc{gmm} and \textsc{km} classifications of agents as authentic
or inauthentic directly provide \emph{jury selection procedures} (\textsc{jsp}s):
select the largest jury that includes only agents classified as authentic.
This defines the \textsc{gmm} and \textsc{km} \textsc{jsp}s.

\subsubsection*{Evaluation Conditions}

To evaluate the \textsc{gmm} and \textsc{km} \textsc{jsp}s, we compare
the majority correctness scores of the juries they select from $\mathcal{P}_{\mathtt{All}},\mathcal{P}_{\mathtt{B}_{\uparrow}},\mathcal{P}_{\mathtt{D}_{\uparrow}}$
and $\mathcal{P}_{\mathtt{L}_{\uparrow}}$ with $r=500$. The low
number of authentic agents and rounds result in more diffuse clustering
environments and situations in which inauthentic agents have strong
negative effects on \textsc{mcs}s (cf. Sec. \ref{subsec:Baseline-Majority-Correctness}).

\subsubsection*{Expected Juries}

For each \textsc{jsp}, population, and noise level, we produce $3$
expected juries—the average, best and worst cases—based on mean misclassification
scores and standard deviations. Let $\mathcal{P}:A\longrightarrow\{\mathtt{A},\mathtt{B}_{i},\mathtt{D}_{i},\mathtt{L}_{i}\}_{i\in\{\uparrow,\downarrow,\updownarrow\}}$
be a population. Assume that for each agent type $\mathtt{X}\in\{\mathtt{A},\mathtt{B}_{i},\mathtt{D}_{i},\mathtt{L}_{i}\}_{i\in\{\uparrow,\downarrow,\updownarrow\}}$,
we have a misclassification score $\delta_{\mathtt{X}}$. As a \textsc{jsp}
removes agents classified as inauthentic, the selected jury will contain
$(1-\delta_{\mathtt{A}})|\mathtt{A}|$ authentic agents and $\delta_{\mathtt{Y}}|\mathtt{Y}|$
inauthentic agents, for each inauthentic agent type $\mathtt{Y}\in\{\mathtt{B}_{i},\mathtt{D}_{i},\mathtt{L}_{i}\}_{i\in\{\uparrow,\downarrow,\updownarrow\}}$.
As specific agent identity does not matter for behavior and thus for
\textsc{mcs}s, it is not important exactly \emph{which} agents of
each type are removed, only the percentage is important. We implemented
selecting agents as juries as follows:

For each agent type $\mathtt{X}\in\{\mathtt{A},\mathtt{B}_{i},\mathtt{D}_{i},\mathtt{L}_{i}\}_{i\in\{\uparrow,\downarrow,\updownarrow\}}$,
let $\mathtt{X}^{-1}$ denote the agents of that type, i.e., $\mathcal{P}^{-1}(\mathtt{X})\subseteq A$.
Enumerate each such $\mathtt{X}^{-1}$ so that $\mathtt{X}^{-1}=\{x_{1},...,x_{|\mathtt{X}^{-1}|}\}$.
Let $\delta_{\mathtt{X}}$ represent a misclassification score, and
let $\lfloor\cdot\rfloor$ be the floor function which rounds down
to closest integer. Given this, we define the agents of type $\mathtt{X}$
to \emph{keep} in the jury to be 
\[
\mathtt{X}(\delta_{\mathtt{X}})=\{x_{k}\in\mathtt{X}^{-1}\colon k\leq\min(\lfloor|\mathtt{X}^{-1}|\cdot\delta_{\mathtt{X}}\rfloor,|\mathtt{X}^{-1}|)\}.
\]
I.e., $\mathtt{X}(\delta_{\mathtt{X}})$ contains the first $(1-\delta_{\mathtt{X}})$
percent (rounded down to closest integer) of $\mathtt{X}^{-1}$. The
$\min$ operation is needed as $\delta_{\mathtt{X}}$ may exceed $1$
if standard deviation is added.

To obtain the average, best and worst case expected juries, we either
directly use $\delta_{\mathtt{X}}$ equal to the mean misclassification
score $M_{\mathtt{X}}$ of $\mathtt{X}$, or add or deduct two standard
deviations. I.e., we use $\delta_{\mathtt{X}}=M_{\mathtt{X}}$ for
the average, $\delta_{\mathtt{X}}=\max(0,M_{\mathtt{X}}-2\sigma_{\mathtt{X}})$
for the best, and $\delta_{\mathtt{X}}=\min(M_{\mathtt{X}}+2\sigma_{\mathtt{X}},1)$
for the worst case jury. The average {[}best / worst{]} expected jury
of $\mathcal{P}$ is then the set of agents
\[
{\textstyle \mathtt{G}(1-\delta_{\mathtt{G}})\cup\bigcup_{\mathtt{Y}\in\{\mathtt{A}_{i},\mathtt{D}_{i},\mathtt{L}_{i}\}_{i\in\{\uparrow,\downarrow,\updownarrow\}}}\mathtt{Y}(\delta_{\mathtt{Y}})}
\]
with $\delta_{\mathtt{X}}=M_{\mathtt{X}}$ {[}$\delta_{\mathtt{X}}=\max(0,M_{\mathtt{X}}-2\sigma_{\mathtt{X}})$
/ $\delta_{\mathtt{X}}=\min(M_{\mathtt{X}}+2\sigma_{\mathtt{X}},1)${]}
for each $\mathtt{X}\in\{\mathtt{A},\mathtt{B}_{i},\mathtt{D}_{i},\mathtt{L}_{i}\}_{i\in\{\uparrow,\downarrow,\updownarrow\}}$.

I.e., the average expected jury contains the mean percentage of authentic
agents classified as authentic plus the mean percentage of each inauthentic
type misclassified as authentic. The best and worst cases are similar,
just factoring in standard deviation.

\subsubsection*{Jury Results}

Table~\ref{tab:MCS:E+F_results} summarizes \textsc{gmm} and \textsc{km}
\textsc{jsp}'s mean majority correctness scores and SD for populations
$\mathcal{P}_{\mathtt{B}_{\uparrow}},\mathcal{P}_{\mathtt{D}_{\uparrow}},\mathcal{P}_{\mathtt{L}_{\uparrow}}$
and $\mathcal{P}_{\mathtt{All}}$ in key conditions. In $\mathcal{P}_{\mathtt{B}_{\uparrow}}$,
\textsc{km }and \textsc{gmm} \textsc{jsp}s\textsc{ }result in \textsc{mcs}s
that clearly show how removing agents classified as inauthentic from
the baseline jury suffices to yield perfect \textsc{mcs}s, despite
13\% misclassification among inauthentic agents by both \textsc{km}
and \textsc{gmm} (Table~\ref{tableD}). Similar observations hold
for $\mathcal{P}_{\mathtt{D}_{\uparrow}}$, where the \textsc{gmm}
\textsc{jsp} achieves maximum \textsc{mcs}. The setback in the \textsc{km}
high noise case is explained by difficulties in distinguishing authentic
from inauthentic agents: The non-precautious misclassification of
inauthentic agents as authentic forecloses the jury to achieve a higher
\textsc{mcs} when the inauthentic agents are activated. For \textsc{gmm
}and\textsc{ km jsp}s in $\mathcal{P}_{\mathtt{L}_{\uparrow}}$, we
might expect lower \textsc{mcs}s given rather substantial misclassification
numbers in the high noise case (Table~\ref{tableD}). Yet, we observe
perfect \textsc{mcs}s, explained by the non-coordinated way that $\mathtt{L}_{\uparrow}$
act inauthentically. Hence, difficulties classifying this subgroup
for both classifiers are mitigated by its limited effect on lowering
\textsc{mcs}s. Note how $\mathcal{P}_{\mathtt{L}_{\uparrow}}$ is
unaffected by the filter in Table~\ref{tableD}: $\mathtt{L}_{\uparrow}$
agents act on their individual beliefs about quality, i.e., they act
uncoordinated on their personal property, which cannot be filtered
for per voting round without changing the population size.

Assessing \textsc{jsp}s given $\mathcal{P}_{\mathtt{All}}$, we show
the \textsc{mcs}s achieved through the \textsc{gmm} method strictly
dominate those from \textsc{km} in low and mid noise cases, both in
terms of mean value and SD. Moreover, the \textsc{gmm jsp} strictly
outperforms baseline juries in all noise cases when looking at average
and best juries. Merely in high noise, worst case, we observe that
neither the\textsc{ gmm} nor the \textsc{km jsp} outperforms the baseline
jury.

\section{\label{sec:Concluding-Remarks}Concluding Remarks}

Influence or information operations such as coordinated inauthentic
behavior (CIB), e.g. performed by attention hacking bots, shape public
opinion by elevating or suppressing topics through coordinatedly up-
or downvoting social media posts, mimicking authentic behavior to
avoid detection, nullifying online voting judgments' reliability.
To restore wisdom-of-crowds effects, this paper designed two accurate
and feasible \emph{jury selection procedures} (\textsc{jsp}s) that
discard agents classified as inauthentic from the voting jury.

Comparing the \textsc{gmm }and\textsc{ km jsp}s, the main difference
is accuracy: The \textsc{gmm} \textsc{jsp} detects more inauthentic
agents, exhibiting smaller false negative errors and hence more precaution.
Both \textsc{jsp}s select juries with vastly increased \emph{majority
correctness scores} (\textsc{mcs}s), with preponderantly better scores
for the \textsc{gmm jsp}. Overall, the application of either almost
fully restores wisdom-of-crowds effects, despite the presence of inauthentic
agents. In the low and mid noise cases, inauthentic agents strongly
affect the baseline \textsc{mcs}s negatively, but both \textsc{jsp}s
successfully eliminate this effect. Only in populations with a high
degree of hiding (i.e., high noise, where inauthentic agents act mainly
authentically), the \textsc{jsp}s do not significantly increase \textsc{mcs}s.
However, in these cases the inauthentic agents also exhibit negligible
negative effects on \textsc{mcs}s.

The latter highlights a trade-off for inauthentic attention hacking
behavior: attention hackers must balance their accounts' activity
to, on the one hand, hide their true identity by acting authentically,
and, on the other, act in a coordinated manner to sway the majority
vote. We believe this may be exploited in designing attention hack
resistant social media vote systems. Employing \textsc{jsp}s means
inauthentic actors must hide more often, raising the cost of influence
for the attention hackers that handle them. Further, \textsc{jsp}s
could be combined with a user reputation system that only publicly
displays a user's vote if the user has logged enough (ignored) votes.
Beyond raising bot startup costs, this may provide early data for
\textsc{jsp}s.\medskip{}

\noindent We round off with a discussion of ethical considerations,
model assumptions, and data collection.

\subsection{Ethical Considerations}

Any suppression of information in public fora raises ethical concerns
about censorship. The suppression of reactions to social media posts
is no different. Generally, we find that the suppression of coordinated
inauthentic behavior as used by attention hackers is defendable, justified
by the aim to combat misinformation online. We omit further discussion
of this point. However, in applying automated techniques based on
classification, there is always a risk that misclassification occurs.
If the classification is used for censorship—as is the case here—misclassification
may then lead to\emph{ }unrightful censorship.

The \textsc{jsp}s risk unjustified censorship on two points: the unrightful
censorship of individuals due to behavioral correlation with inauthentic
agents, and the unrightful censorship of groups due to an authentic
disagreement with the notion of quality assumed by \textsc{jsp} deployers.

Concerning individuals, then we designed the \textsc{jsp}s with a
focus on the two stated desiderata \emph{vox populi} (to minimize
false positive errors, i.e., to preserve as many authentic agents
as possible) and \emph{precaution} (to minimize false negative errors,
i.e., to eliminate as many inauthentic agents as possible). Vox populi
implies a desire to not unrightfully censor individuals, but is opposed
by precaution: the most precautious model censors all, while the model
that preserves most voices censors none. Given our ABM and its parameters,
employing \emph{ends-justify-the-means }reasoning, and taking the
correct evaluation of posts' quality to be the primary end, we find
it worth compromising vox populi over deprioritizing precaution: as
illustrated in Figures~\ref{fig:plot-A} and \ref{fig:plot-B}, deprioritizing
precaution quickly threaten the wisdom-of-crowds effect as few inauthentic
agents in the jury drastically lower the majority correctness score,
while compromising with vox populi by allowing small fractions of
authentic agents to be labelled as inauthentic is—with respect to
\textsc{mcs}—absorbed by the wisdom of crowds exhibited by even a
small jury of only authentic agents.

In the classification, the balance between vox populi and precaution
is controlled by the classification threshold. As classification threshold,
we precautiously chose that agents should be labelled authentic $4$
of $5$ times to be classified as authentic. This choice did not cause
tremendous collateral damage to vox populi. While we deem especially
the \textsc{gmm jsp} a precautious method, it still exhibits low ($<.11$)
false positive misclassification errors throughout, except for $\mathcal{P}_{\mathtt{All}}$
in high noise. The \textsc{km jsp}, similarly shows low false positive
errors $(<.1)$ except for $\mathcal{P}_{\mathtt{All}}$ in low and
high noise (cf. Table~\ref{tableD}). The approach remains flexible
to emphasizing vox populi further by lowering the $\nicefrac{4}{5}$
classification threshold.

Concerning group censorship, it is relevant that our approach assumes
an agreed-upon notion of \emph{truth about the quality of posts} for
which a commonly acknowledged arbiter exists. This is a fundamental
premise of our method: if no such notion exists, majority correctness
scores loose their meaning and the assumptions of the classifiers
are unmet. Such a notion of quality is of paramount importance in
relation to fake news, where, arguably, ``objective'' quality exists,
embodied e.g. by the Principles of Journalism. However, the criteria
for what constitutes quality may lead to marginalization of groups.
E.g., sympathizers of Alex Jones and InfoWars might be marginalized
by censorship if quality is equated with adhering to the Principles
of Journalism, or sympathizers of the black feminist Combahee River
Collective may be marginalized if quality is equated with adhering
to ideals of the National Association for the Advancement of Colored
People of the 1970s. Therefore, the notion of quality used in applications
should be carefully defined, and preferably made open to the public
e.g. by inclusion in community standards or terms and conditions of
social media platforms.

Due to the risk of unrightful censorship, we would always suggest
that users are made aware of censorship decisions that concern them
and are given the option to appeal. This, of course, also allows accounts
used in IOs to appeal, but appeal adds a non-trivial maintenance cost
to e.g. large bot collectives.

\subsection{\label{subsec:Assumptions}Assumptions of the ABM and Classification}

While our contribution hopefully serves as a proof of concept for
jury selection procedures as a tool to counter reaction-oriented CIB-based
IOs, the simulated environment is not in a one-to-one correspondence
with the plethora of environments found on social media platforms.
We discuss how modeling choices relate to social media platforms and
how assumptions may be relaxed, first concerning the ABM, then the
classification.

On social media platforms, it is likely that human users at times
vote inauthentically to a low degree that should not be penalized
by censorship. Such inauthentic voting violates the ABM's assumptions
about authentic agents who vote given only their competence-based
beliefs. Our classification results indicate, however, that the authenticity
assumption may be relaxed. \emph{Lone wolfs} in the high noise case
behave \emph{almost} authentically, and may be interpreted as generally,
but not fully, authentic, uncoordinated users. These agents are further—by
the \textsc{gmm jsp}—often \emph{misclassified }as authentic in high
noise (cf. $\mathtt{L}$ sections of Figure~\ref{fig:figureC} and
Table~\ref{tableD}), but correctly classified for low and mid noise,
which indicates that the \textsc{gmm} method may be tuned to tolerate
a degree of uncoordinated, inauthentic behavior.

Further, on social media, vote participation is not complete: most
users do not react to most posts. For simplicity, we have not included
abstaining as an option in the ABM, but all steps including \textsc{mcs}
calculation and jury selection would be unaffected. As we return to
below, also the classification can accommodate for a less complete
vote participation.

Concerning classification, disciplines not directly related to social
media applications and misinformation research show how dimensionality
reduction and SVD procedures can be applied to empirical data to disclose
coordinated voting groups and patterns: US Congress roll call votes
have been clustered based on scores similar to the weighted eigenvectors
used in this paper \citep{Yang2020_1dim_congress,Porter2005,Sirovich2003,Poole2000}.
SVD Scatterplots of votes as suggested by \citet{Porter2005}, for
instance, provide proxies for party stance. While \citet{Yang2020_1dim_congress}
explore roll call vote data only $1$-dimensionally, we expand the
application and cluster on $2$ partial components; both their and
our applications can be generalized to more dimensions to increase
precision in less exposing vote environments. Moving towards social
media applications, this can become relevant for votes with not only
binary but several options from which to choose, such as vote data
reflecting Facebook's $6$ reactions.

We rely on unsupervised methods that disclose coordination that go
unnoticed by supervised methods that take only features of individual
accounts into consideration \citep{Khaund2022_Socialbotcoord,Orabi2020,Grimme2018CoordUnsupervOwnBots,cresci2017paradigm}.
We add a single supervised learning element—logistic regression—to
apply labels to agent clusters found by the unsupervised steps. In
the logistic regression, we have used that authentic votes correlate
with post quality (possibly allowing for noise in observing quality).
Other subjective assumptions could be used to steer labeling while
producing equally efficient jury selection procedures.

Besides limiting supervision, the input data needs of the \textsc{gmm}
and \textsc{km }jury selection procedures are vastly more feasible
than \citet{GaleazziRendsvigSlavkovik2019}'s: we rely on $500$ observations,
where the $\chi^{2}$ test would require at least $2^{1000}$ for
our population $\mathcal{P}_{\mathtt{All}}$. In empirical application,
obtaining $500$ votes of one user group may still be a challenge.
A mitigating factor is that the proposed \textsc{jsp}s can accommodate
for missing data, and, for validation, only the inauthentic agents
need to be fixed over several voting rounds, while the authentic agents
may vary, as these vote independently. Thus, we can lift the assumption
that all agents are always presented with, and vote, on every post.

\subsection{Empirical Validation and The Release of Reactions Data}

Empirical data—in contrast to simulated data—to further validate jury
selection procedures remains difficult to obtain \citep{Disinfresearch-agenda-2020,misinfo_data,Torres-Lugo_Likes_Manip_deletions}.
Among the platforms that provide APIs for academic purposes, only
Twitter releases user-IDs of (public) profiles that have clicked the
like-button. However, while Twitter provides generous academic access
to historical data for researchers, the platform does not allow to
automatically scrape \emph{comprehensive} lists of users that have
liked, but only releases the user-IDs of the 100 \emph{most recent}
liking users of any single post. Additionally, lists of liking users
may be requested at most 75 times per 15 minutes. For small-scale
Twitter environments where posts receive few likes, these restrictions
may be balanced by using a suitably timed algorithm. However, for
large political hashtags like \texttt{\#MakeAmericaGreatAgain} or
\texttt{\#Brexit} where CIB-based IOs may be feared to be in play,
the current data restrictions make it practically impossible to obtain
a complete picture of liking behavior.

The proof of concept for \textsc{jsp}s provided in this paper provides
a direct use case for reactions data in the fight against online misinformation.
The data is necessary to evaluate, tweak and deploy the suggested
methods. The paper thus provides a direct argument for a more comprehensive
release of and access to reactions data to researchers, e.g. under
full anonymization and non-disclosure agreements or via open API access
to publicly available data.

\end{document}